\documentclass{iaus}
\usepackage{graphicx,natbib,amsmath,times}
\usepackage{ifpdf}

\title[Modeling of multi-planet extrasolar systems] 
{Stability constraints in modeling of multi-planet extrasolar systems}

\author[K. Go\'zdziewski, C. Migaszewski and A. Musieli\'nski]   
{Krzysztof Go\'zdziewski, Cezary Migaszewski, Arek Musieli\'nski}
\affiliation{Toru\'n Centre for Astronomy, \\ N. Copernicus University,
PL-87-100 Toru\'n, Poland \\ email: {\tt
\{k.gozdziewski,c.migaszewski,a.musielinski\}@astri.umk.pl} 
}
\pubyear{2008}
\volume{249}
\pagerange{0-0}
\jname{Exoplanets: Detection, Formation and Dynamics}
\editors{eds.}
\newcommand\Chi{{(\chi^2_\nu)^{1/2}}}
\def\idm#1{{\mbox{\small #1}}}
\def\hda{HD~73526}
\def\hdb{HD~155358}
\def\vec#1{{\boldsymbol #1}}
\begin{document}
\maketitle
%
\begin{abstract}
We present an analysis of high precision radial velocity (RV) observations of stars
hosting multi-planet systems with Jovian companions. We use dynamical
stability constraints and quasi-global methods of optimization. As an
illustration, we present new results derived for the RV data of the Sun-like dwarfs
HD~155358 and $\tau^1$~Gruis.
\end{abstract}

\section{Introduction}
%
Extrasolar planetary systems have became a major challenge for contemporary
astrophysics and dynamical astronomy.  One of the most difficult problems in
this field concerns the orbital stability of such systems, in particular when
related to the observations and their interpretation.   Usually, the
investigations of long-term evolution are the domain of direct, numerical
integrations. The stability of planetary systems is often understood in terms of
the Lagrange definition implying that orbits remain well bounded over infinite
time. Other definitions may be formulated as well, like the astronomical
stability \citep{Lissauer1999} requiring that the system persists over a very
long, Gyr time-scale or Hill stability \citep{Szebehely1984} that requires the
constant ordering of the planets. In our studies, we prefer a more formal and
stringent definition related to the fundamental Kolmogorov-Arnold-Moser Theorem
(KAM), see \cite{Arnold1978}.  Planetary $N$-body systems, involving a dominant
mass of the parent star and significantly smaller planetary masses, are well
modeled by close-to-integrable, Hamiltonian dynamical systems. According to the
KAM Theorem, their evolution may be quasi-periodic (with a discrete number of
fundamental frequencies, which are stable forever --- and also stable  with
respect to the other notions of stability quoted above), periodic (or resonant;
stable or unstable) or chaotic (with a continuous spectrum of frequencies, and
unstable). In the last case, initially close phase trajectories diverge
exponentially, i.e., their Maximum Lyapunov Characteristic Exponent (MLCE,
denoted also by $\sigma$) is positive.   This understanding of the global
structure of the phase space is widely adopted, in particular with regard to the
Solar system dynamics
\cite[e.g.,][and references therein.]{Wisdom1991,Laskar1990,
Holman1996,Malhotra1988,Nesvorny1999,Robutel2001,
Murray2001,Michtchenko2001,Lecar2001,Morbidelli2002}. 
However, the distinction between regular and chaotic trajectories is a difficult
task that, in practice, may be resolved only with numerical methods relying on
efficient and accurate integrators of the equations of motion.

\section{Stability indicators}
%
To detect chaotic motions in the phase space, many numerical tools are
available. Concerning the dynamics of close to integrable Hamiltonian systems,
these tools can be roughly divided onto two classes: spectral algorithms that
resolve the fundamental frequencies and/or their diffusion rates
\citep{Laskar1993,Nesvorny1997,Michtchenko2001}, and methods based on the
divergence rate of initially close phase trajectories, expressed in terms of the
Lyapunov exponents \citep{Benettin1980,Froeschle1984}.  These fast indicators
can be correlated with geometrical evolution of orbital osculating elements,
like the maximal eccentricity ($\max e$), maximal amplitude of the critical
angle of a resonance ($\max \theta$) or with event time $T_E$ (indicating a
collision or an ejection of a body from the system). Unfortunately, no general
relation between these indicators can be determined
\citep{Lecar2001,Michtchenko2001}. The event time $T_E$,   relying on
CPU-intensive long-term integrations of the equations of motion 
\citep{Lecar2001} can be considered as a direct measure of the  astronomical
stability.

In our work, among the the spectral tools, we often choose the method invented
by \citet{Michtchenko2001}; its idea is very simple --- to detect chaotic
behavior one counts the number of frequencies in the FFT-spectrum of an
appropriately chosen dynamical signal. We deal with conservative Hamiltonian
systems; so in a regular case, the spectrum of fundamental frequencies is
discrete and we obtain only a few dominant peaks in the FFT spectrum. Chaotic
signals do not have well defined frequencies, and their FFT spectrum is very
complex. The number of peaks in the spectrum above some noise level $p$
(typically, $p$ is set to a few percent of the dominant amplitude) tell us about
the character (regular vs chaotic) of a given phase trajectory.

The basic tool to discover exponentially unstable bounded orbits, i.e. chaotic
orbits, is the Maximum Lyapunov Characteristic Exponent (MLCE) $\sigma$.  The
direct computation of the MLCE is based on the analysis of the tangent vectors
$\vec{\delta}$ which are solutions to the variational equations of the equations
of planetary motions:
\[
\frac{\mbox{d}}{\mbox{d}t} \vec{x} =  \vec{f}(\vec{x}),
\]
where $\vec{x}$ denotes the state vector including coordinates and momenta, and
$\vec{f}$ stands for the gravitational forces. For its solution
$\vec{\phi}=\vec{\phi}(t)$ we define:
\begin{equation}\label{eq}
\frac{\mbox{d}}{\mbox{d}t} \vec{\delta} =  \vec{A}(t) \vec{\delta},
\quad \vec{A}(t) := \frac{\partial \vec{f}}{\partial \vec{x}}[\vec{\phi}(t)],
\quad \delta = ||\vec{\delta}||.
\end{equation}
Asymptotically, the MLCE value is given by \citep{Cincotta2000}:
\begin{equation} \label{sig}
    \sigma = \lim_{t \rightarrow \infty} \frac{1}{t} \int_0^t
    \frac{\dot{{\delta}}(s)}{{\delta}(s)} \mbox{d}s.
\end{equation}
If $\sigma$ converges to some positive value, we conclude that the nominal orbit
$\vec{\phi}$ and some initially close orbit diverge exponentially at the rate
$\exp(\sigma t)$. Two practical difficulties arise when the direct definition
(\ref{sig}) is used: the convergence of $\sigma$ is often very slow, and it is
difficult to tell how small the final value of $\sigma$ should be to consider it
$\sigma=0$.

A large variety of methods has been proposed to overcome the problem of slowly
convergent MLCE estimates. Recently,  a new algorithm offering excellent
convergence, called MEGNO (Mean Exponential Growth factor of Nearby Orbits), has
been  proposed by \citet{Cincotta2000}. The definition of MEGNO and its mean
value is the following \citep{Cincotta2003}:
\begin{equation}\label{mfm2}
    Y(t) = \frac{2}{t} \int_0^t
    \frac{\dot{{\delta}}(s)}{{\delta}(s)} s \mbox{d}s , \qquad
    \left<Y\right>(t) = \frac{1}{t} \int_0^t Y(s) \mbox{d}s.
\end{equation}
It was shown that  if $\vec{\phi}(t)$ is a regular solution with a linear
divergence of nearby orbits then $\lim_{t\rightarrow\infty} \left<Y\right>(t) =
2$, and if $\vec{\phi}(t)$ is a chaotic  solution then $\left<Y\right>(t) \sim
(\sigma/2) t$, as $t \rightarrow \infty$. Moreover, when $\left<Y\right>(t)$
tends towards a  value different from 2, then it indicates that close
trajectories diverge according to a certain power law. If $\vec{\phi}(t)$  is  a
periodic solution then $\left<Y\right>(t)$ tends to 0. The asymptotic behavior
of $\left<Y\right>(t)$  is given by a uniform formula $ \left<Y\right>(t) \sim
a t + d, $ where $a \sim 0$ and $d\sim 2$ for a quasi-periodic solution, while 
$a \sim \sigma/2$ and $d\sim 0$ for an irregular and stochastic motion. Having 
$Y(t)$ we can indirectly estimate the MLCE on a finite time interval. The weight
function $s$ in the definition of MEGNO reduces the contribution of the initial
part of the tangent vector evolution, when the exponential divergence is too 
small to be observed relative to other linear and nonlinear effects
\citep{Morbidelli2002}. Thus, fitting the straight line to the final part of
$Y(t)$, we obtain good estimates of $\sigma$ from a relatively shorter piece of
trajectory than in the direct MLCE evaluation.

\section{Modeling the RV data -- an overview}
%
The radial velocity (RV) is still the most efficient technique for detecting
extrasolar planets. To model the RV signal, the standard formulae
by~\cite{Smart1949} are commonly used. Each planet in the system contributes to
the reflex motion of the star at time $t$ with:   
\begin{equation}
V_{\idm{r}}(t) = K [ \cos (\omega+\nu(t))  + e \cos \omega] + V_0,
\label{eq:eq1}   
\end{equation} 
where $K$ is the semi-amplitude, $\omega$ is the argument of pericenter,
$\nu(t)$ is the true anomaly (involving implicit dependence on the orbital
period $P$ and time of periastron passage $T_{\idm{p}}$), $e$ is the
eccentricity, $V_0$ is the velocity offset. We interpret the primary model
parameters $(K,P,e,\omega,T_{\idm{p}})$ in terms of the Keplerian elements and
minimal masses related to coordinates of Jacobi \citep{Lee2003} or Poincar\'e
\citep{FerrazMello2006}.

In our previous work, we tested and tried to optimize different tools helpful
for exploring the multi-parameter space of $\Chi$ for the model Eq.~\ref{eq:eq1} and
its generalizations.  In the case when $\Chi$ may possess many local extrema, we
found that  good results can be obtained with hybrid optimization
\citep{Gozdziewski2006b}. A single  run of  the hybrid code starts the 
quasi-global genetic algorithm \citep[GA,][]{Charbonneau1995}. GA 
makes it easy to carry out
a constrained optimization within prescribed parameter bounds or to add
a penalty term to $\Chi$. The best fits found with GAs are not very accurate in
terms of $\Chi$, so finally, a number of the best fit members of the
``population''  are refined using a relatively fast local method like the simplex
of Melder and Nead~\citep{Press1992}.  The simplex is a matter of choice, so we
could use other fast local methods. However, the code using non-gradient methods
works with minimal requirements for user-supplied information. It is
only required to define the model function ]the so called {\em fitness function}, usually
equal to $1/\Chi$] --- conveniently, this function is the same for the GAs and simplex --- and
to determine (even very roughly) the bounds of the parameters.  The  repeated runs
provide an ensemble of the best-fits that helps us to detect local minima of
$\Chi$, even if they are distant in the parameter space. We can also obtain
reliable approximation to the parameter errors  \citep{Bevington2003} within the
$1\sigma$, $2\sigma$ and $3\sigma$  confidence intervals of $\chi^2$ at selected
2-dim parameter planes.   

While we prefer the GAs as the quasi-global optimization tool, other efficient
and robust algorithms for identifying and characterizing multiple planetary
orbits in precision RV data are known. In particular, the Bayesian Kepler
periodogram \citep{Gregory2007} and Markov chain Monte Carlo (MCMC) technique 
\citep{Ford2005a} are proven to be robust tools for calculating the model
marginal likelihood which is used to compare the probabilities of models with
different numbers of planets and for investigating the uncertainty of parameters
in the orbital solutions.

Due to the limited time-span of the observations, we often encounter a problem 
that the data only 
partially cover the longest orbital period. In this situation, 
it is possible either that $\Chi$ does
not have a well defined minimum, or that its shape is very ``flat'',
so the confidence
levels may cover large ranges of the fit parameters.  To illustrate {\em the
shape} of $\Chi$ in selected 2-dim parameter planes, we  perform a systematic
scanning of the space of initial conditions with the fast Levenberg-Marquardt
(L-M) algorithm \citep{Press1992}. Usually,  for representing such scans, we
choose the semimajor-axis---eccentricity $(a,e)$ plane of the outermost planet. 
We fix $(a,e)$ and then search for the best fit, initiating the L-M algorithm
with  starting points selected randomly (but within reasonably wide parameter
bounds).  The L-M scheme ensures a rapid convergence.  It is heavily
CPU-consuming and may be effectively applied in low-dimensional problems. In
reward it provides a clear picture of the parameter space. 

In the case of resonant or mutually interacting planets, the problem is even
more complex.  How to interpret the RV measurements in that case often remains
an open and  difficult question. The $N$-planet  configurations are
parameterized by at least $5 N+1$~parameters, even assuming that the system is
coplanar. The RV signal in terms of Eq.~\ref{eq:eq1} is degenerate --- we have
no information on the inclinations of orbits and the true masses of the companions.
Moreover, we do not know {\em a priori} the number of planets  in the system, so
the resolved solutions are often not unique. It is also well known that the
Keplerian (kinematic) model is usually not adequate to properly explain the RV
variability. Instead, a self-consistent $N$-body Newtonian model should be
applied \citep{Rivera2001,Laughlin2001}.  Basically, the effects  of mutual
interactions included in the Newtonian model could make it possible to determine
or estimate the inclinations and masses provided that long enough time-series of
precision data are available. Nevertheless, the measurements can be affected by
many sources of error, like complex systematic instrumental effects, short
time-series of the observations, irregular sampling due to observing conditions,
and stellar noise. Little is known on the real statistical characteristics of
the error distributions and the assumption that these are Gaussian 
distributions is 
not necessarily valid \citep{FerrazMello2005}, see also \citep{Baluev2007}.

All these factors, in particular the unspecified number of planets and
undetermined or weakly constrained parameters, can  (and often do) lead to 
best-fit solutions representing unrealistic, quickly disrupting configurations
\citep{FerrazMello2005,Lee2006,Gozdziewski2006x}. But according to the
Copernican principle, the detection of strongly unstable systems during two
decades of RV observations is not likely, we would rather expect that the
dynamical stability should be preserved over a significant part of the parent star
life-time counted in Gyrs. Stability is therefore a natural requirement of a
model solution consistent with observations. Many authors take it into account
when analyzing the dynamics of the  best-fit configurations with different
constraints, e.g., to mention only a few examples in an endless list of references:
through long-term integrations, requiring astronomical stability
\citep{Lissauer2001,Laughlin2002} or Lagrange and/or Hill stability
\citep{Barnes2007}, also through fast indicators like the diffusion of
characteristic frequencies \citep{Correia2005}, fast Lyapunov indicators
\citep{Dvorak2005}, maximal eccentricity \citep{Vogt2005,Ford2005b}, critical
angles  \citep[][see also paper by Beaug\'e et al. (2008) in this
volume]{Ji2003,Lee2006,Beauge2006}, or $T_E$ determined over short-time scale
(the Systemic project initiated and led by Greg Laughlin, www.oklo.org).
However, a common approach to search for stable  solutions in a neighborhood of
unstable best-fit configuration by trial and error does not
necessarily provide stable fits  that are simultaneously optimal, in terms of
$\Chi$ or an rms. What is even more important, the stability requirement imply
non-continuous and complex structure of the space of initial conditions that
depends on adopted definition of stability.

Hence, as a general way of modeling the observations we propose  to eliminate 
unstable (for instance, strongly chaotic) solutions {\em during} the fitting
procedure. The idea is very simple: we modify the hybrid algorithm by  adding a
suitable penalty term to the function determining the fit quality, e.g., $\Chi$
or rms for unstable solutions. The penalty term must rely on some signature of
the system stability. We found that MEGNO is particularly useful for that
purpose thanks to its rapid convergence and great sensitivity to chaotic
motions. We have shown that on many examples, this method  (called Genetic Algorithm
with MEGNO Penalty, GAMP) is very useful in modeling resonant or
close-to-resonant planetary configurations when even small errors of orbital
phases or other parameters may lead to quick self-destruction of the system. 
Unfortunately, the algorithm cannot give a definite answer when we want to resolve
the $\Chi$ shape in detail or find strictly stable solutions because,  in
practice, the penalty term  can be calculated only a over relatively short period of
time (due to CPU time requirements). This is particularly important for systems
affected by long-term resonances, i.e., configurations with three and more
planets, inclined orbits \citep{Gozdziewski2007b}. Hence, in an additional 
step, we need to 
examine the stability of individual best fits selected with GAMP over a period
of time related to the time-scale of relevant unstable behaviors (resonances).

\section{\hdb{} - a system with two planets} 
%
To illustrate the algorithms and problems discussed above on a new  planetary
system, we consider the RV observations of \hdb{} by \cite{Cochran2007}. The
data  published in this paper  consist of 71 observations and span $\sim 2100$
days. The single-planet  Keplerian model does not fully explain the RV
variability, so the discovery team studied a 2-planet Keplerian model of the RV.
This yields three acceptable fits: a solution which is stable for 100~Myrs with
$P_{\idm{c}}\sim 195$~days, $P_{\idm{c}}\sim 526$~days and minimizes $\Chi$ as
well as  two unstable fits with $P_{\idm{c}} \sim 500$ and 1500~days,
respectively. These fits have very large $e_{\idm{c}}$ leading to
catastrophically unstable configurations \citep{Cochran2007}.

We try to extend the analysis of the RV data by considering an $N$-body
model of the observations and looking more closely at the phase-space structure
of the putative 2-planet system with the help of the dynamical tools described
above. 

\ifpdf
\begin{figure}
\centerline{
\hbox{
  \includegraphics[width=56mm]{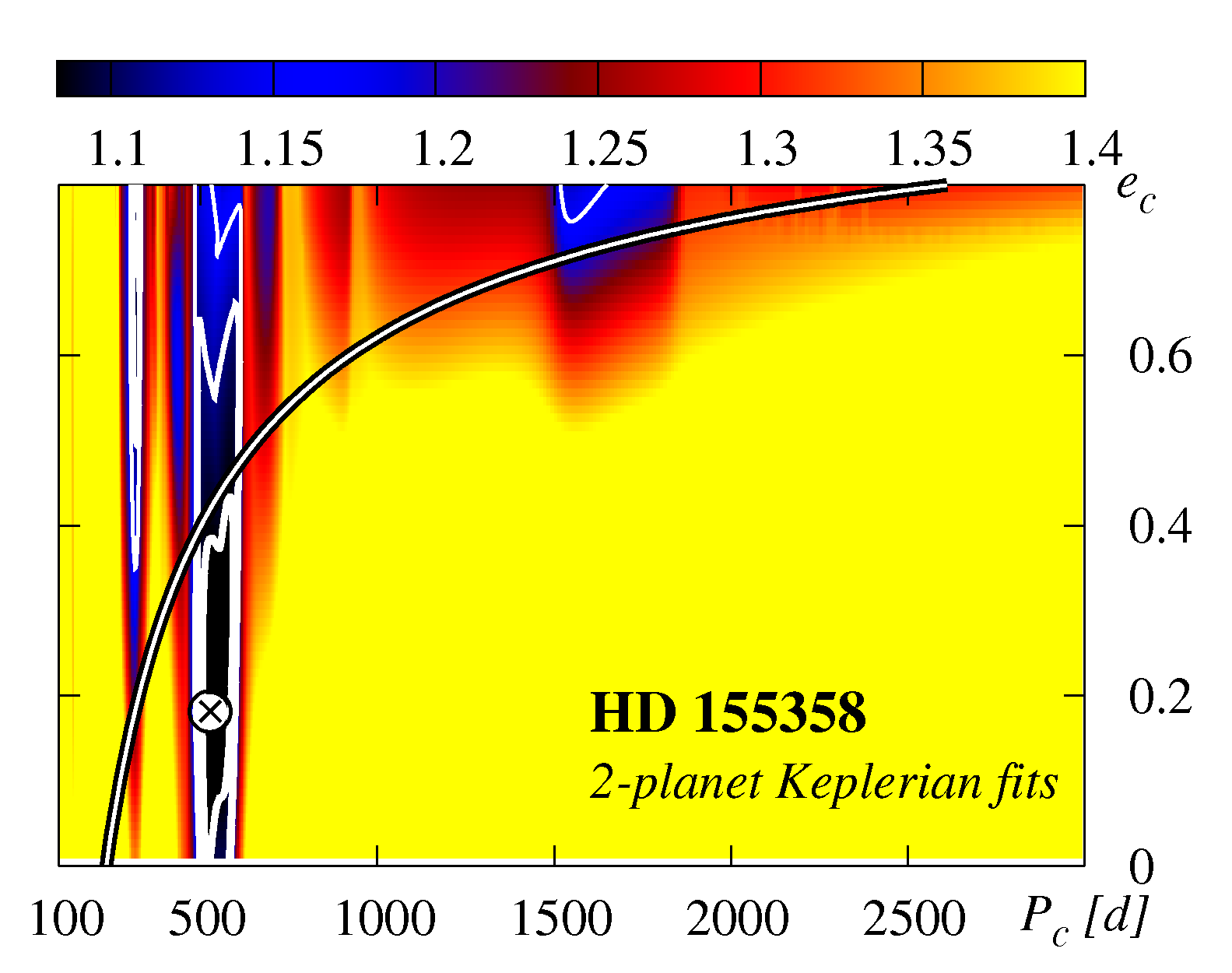}
  \hspace*{1mm}
  \includegraphics[width=56mm]{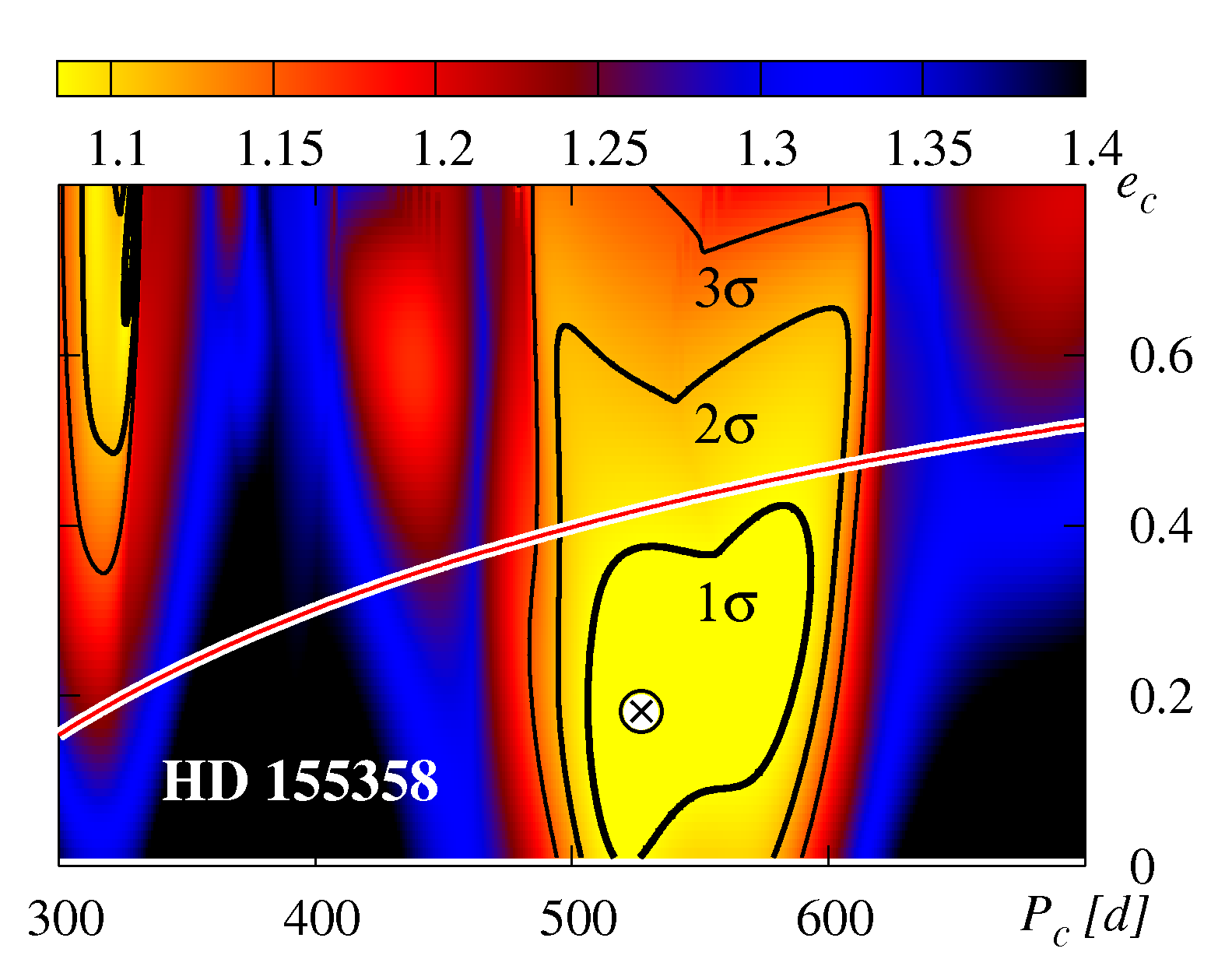}
}
}
\caption{\em 
The levels of reduced $\Chi$ of the 2-planet Keplerian model of the \hdb{} RV
data published in Cochran et al. (2007) onto the
$(P_{\idm{c}},e_{\idm{c}})$-plane. The left panel is for the whole tested range
of parameters. The right panel is for the close-up of the most prominent minima
around  $P_{\idm{c}}\sim 500$~days and $e_{\idm{c}} \sim 0.2$. The smooth line
is for the collision line. Curves labeled with $1\sigma$, $2\sigma$ and
$3\sigma$ are for the confidence levels of the best fit marked with crossed
circle.
}
\label{fig:fig1}
\end{figure}
\else
\begin{figure}
\centerline{
\hbox{
  \includegraphics[width=56mm]{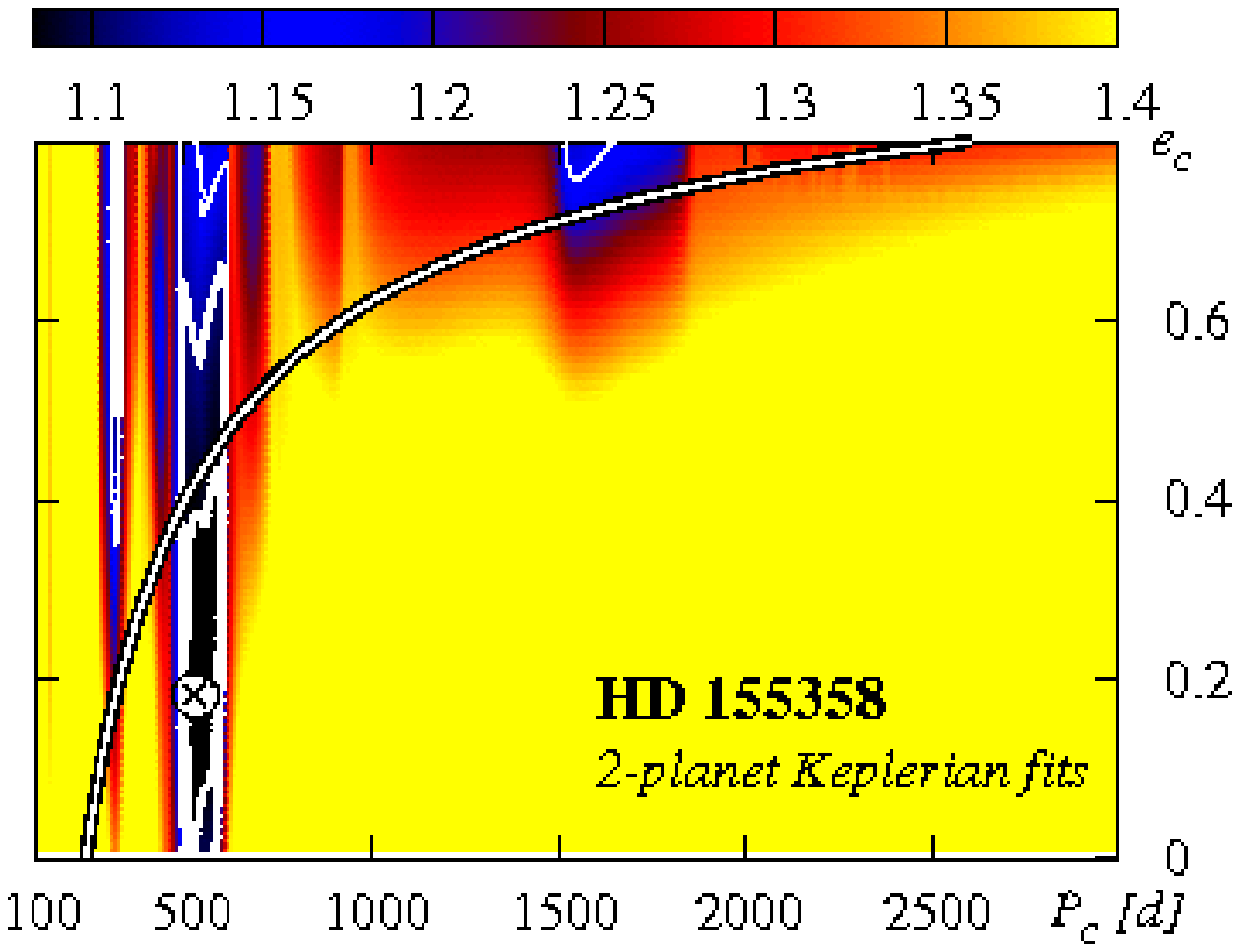}
  \hspace*{1mm}
  \includegraphics[width=56mm]{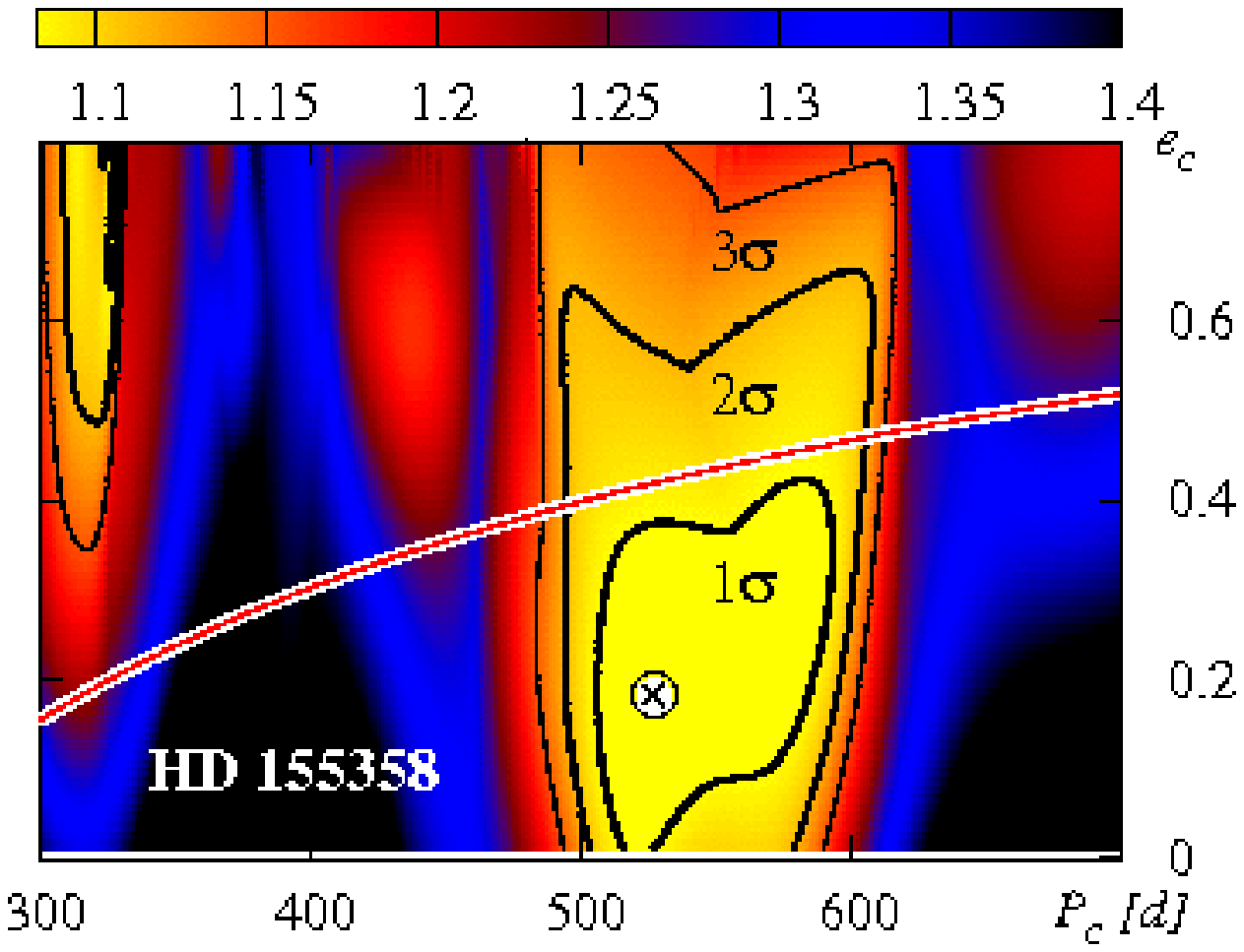}
}
}
\caption{\em 
The levels of reduced $\Chi$ of the 2-planet Keplerian model of the \hdb{} RV
data published in Cochran et al. (2007) onto the
$(P_{\idm{c}},e_{\idm{c}})$-plane. The left panel is for the whole tested range
of parameters. The right panel is for the close-up of the most prominent minima
around  $P_{\idm{c}}\sim 500$~days and $e_{\idm{c}} \sim 0.2$. The smooth line
is for the collision line. Curves labeled with $1\sigma$, $2\sigma$ and
$3\sigma$ are for the confidence levels of the best fit marked with crossed
circle.
}
\label{fig:fig1}
\end{figure}
\fi

At first, we did two systematic scans of $\Chi$ in the
$(P_{\idm{c}},e_{\idm{c}})$-plane. The results are shown in Fig.~\ref{fig:fig1}.
Three local minima of $\Chi$ reported by \cite{Cochran2007} are evident. Two of
them lie far over the collision line of orbits, defined in terms of semi-axes
and eccentricities through $a_b (1+e_b) = a_c (1-e_c)$.  This line marks the zone in
which the mutual interactions of  relatively massive companions can quickly
destabilize the configuration. The solution with $P_{\idm{c}} \sim 300$~days 
yields $\Chi$ similar to that one of the best fit with $P_{\idm{c}} \sim
500$~days. 

Next, we refined the Keplerian fits with $N$-body model. The osculating
parameters of the dominant solution  (see Table~1, fit~II) are slightly
different from that one of the Kepler model,  nevertheless it appears also
stable.  Its neighborhood is illustrated  in dynamical maps shown in
Fig.~\ref{fig:fig2}. Curiously, the best-fit  configuration it located in the
very edge of a stable zone between 5:2 and 3:1~ mean motion resonances (MMRs) of
two planets with masses $\sim 0.8$~m$_{\idm{J}}$ and $\sim 0.5$~m$_{\idm{J}}$,
respectively. As we have  observed in other cases, the $\max e$ and $\max
\theta$ indicators are in excellent  correlation with the measure of formal
stability (here, $\log SN$). We can also see a very complex border of the stable
zone. The system would be located in dynamically active region of the phase
space spanned by a few low-order MMRs. The proximity of the best fit
configuration to the 5:2~MMR and moderate eccentricities indicate a dynamical
similarity of the \hdb{} system to the Solar system.

\ifpdf
\begin{figure}
\centerline{
\hbox{
  \includegraphics[width=52mm]{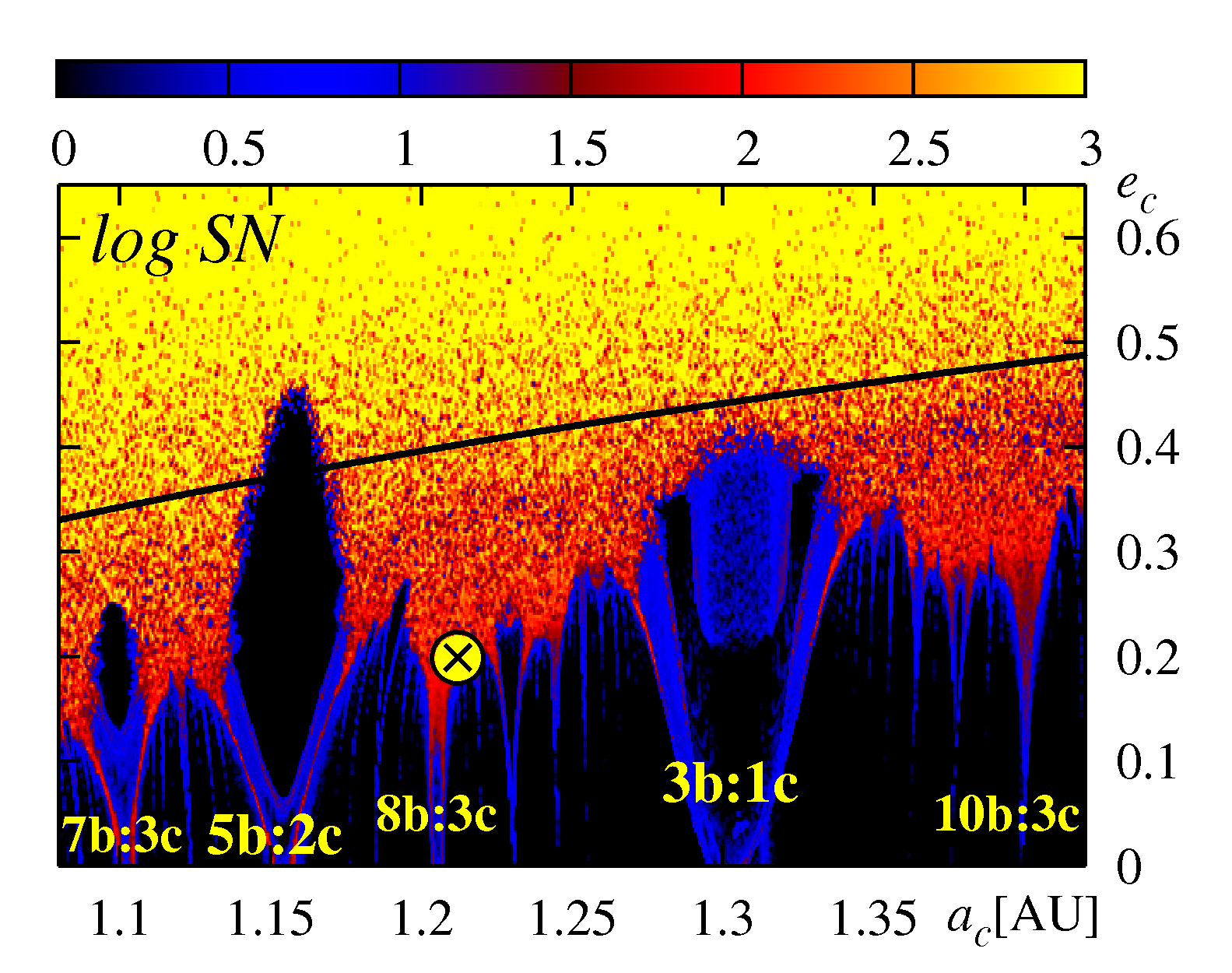}
  \hspace*{6mm}
  \includegraphics[width=52mm]{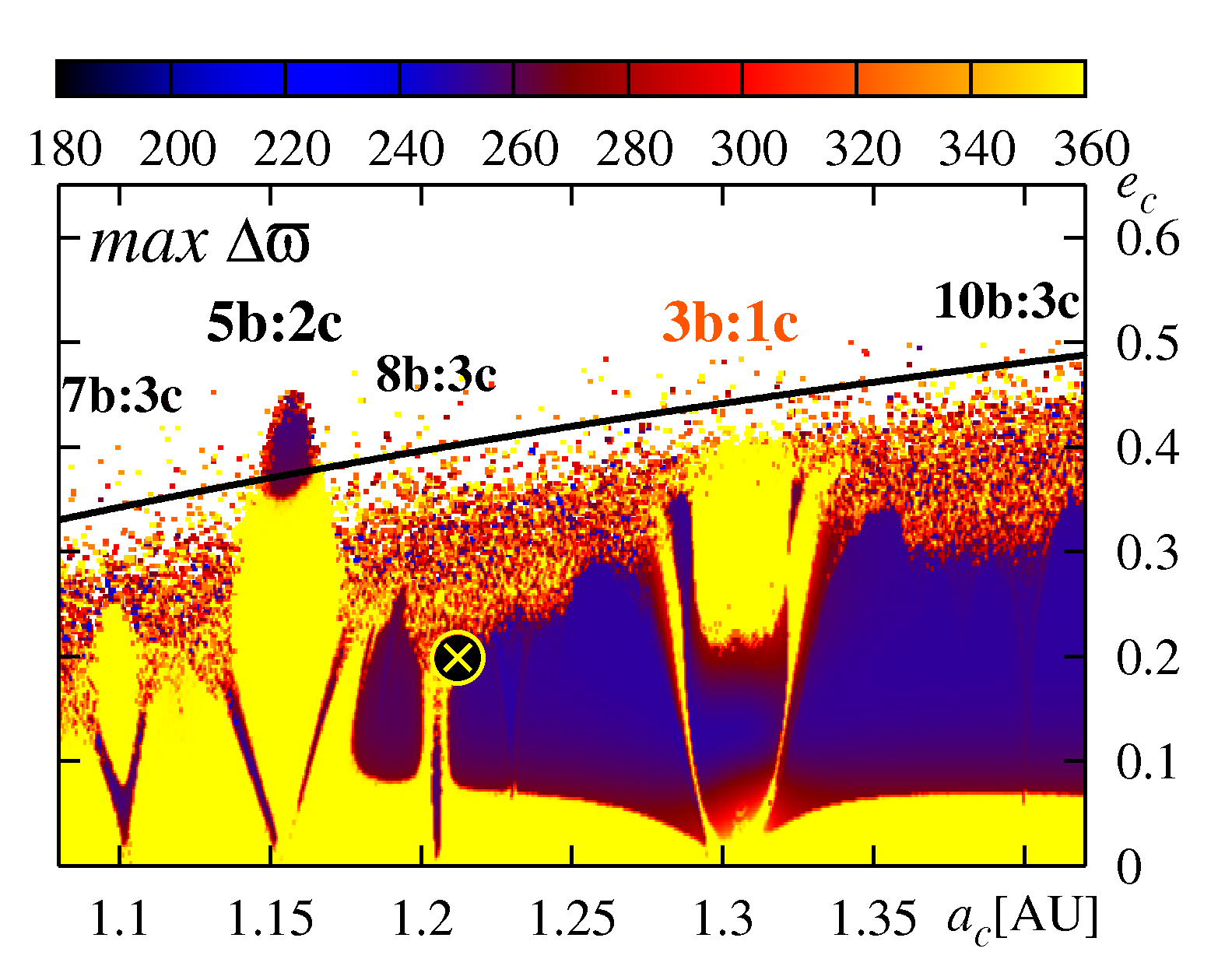}
}
}
\vspace*{0mm}
\centerline{
\hbox{
  \includegraphics[width=52mm]{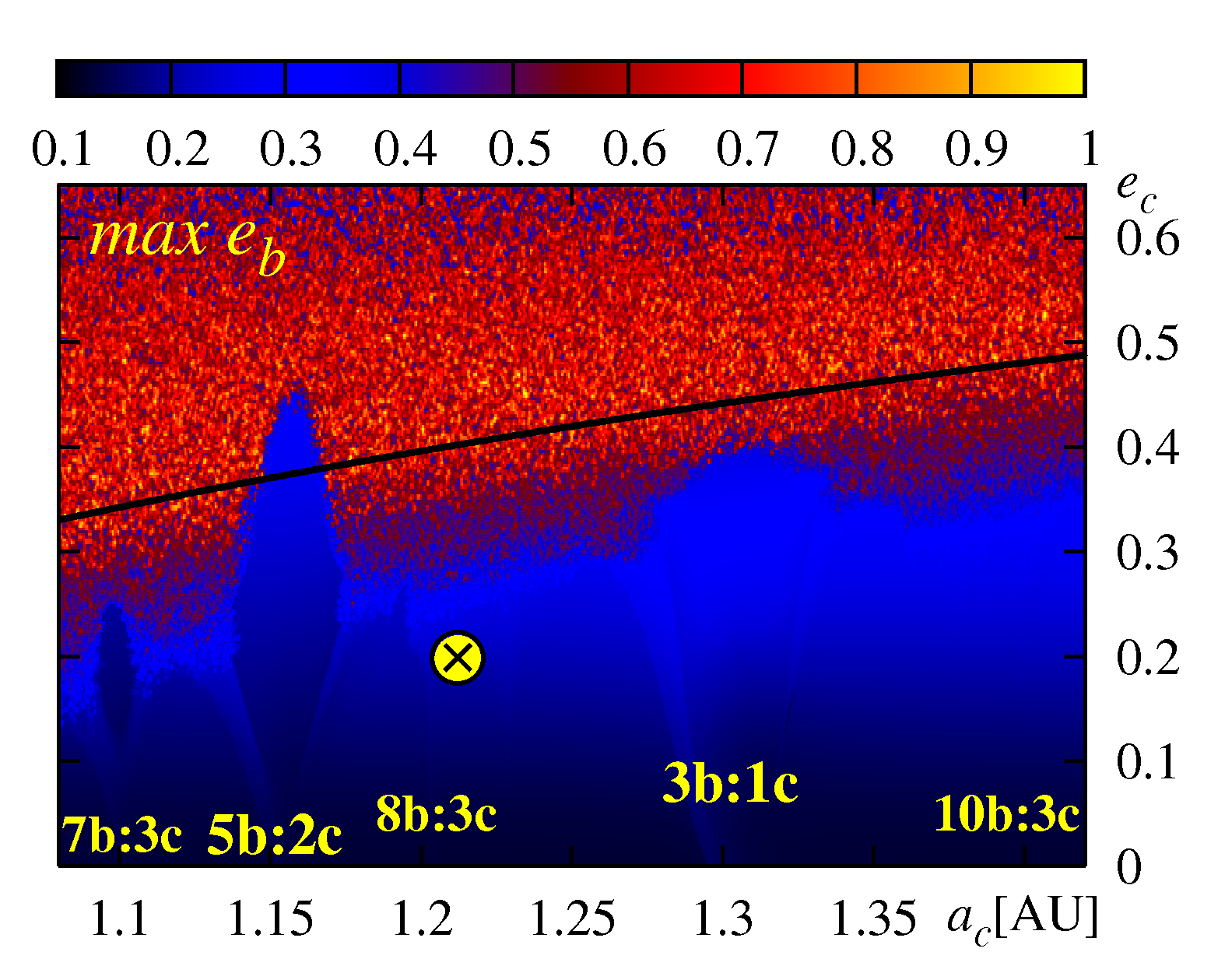}
   \hspace*{6mm}
  \includegraphics[width=52mm]{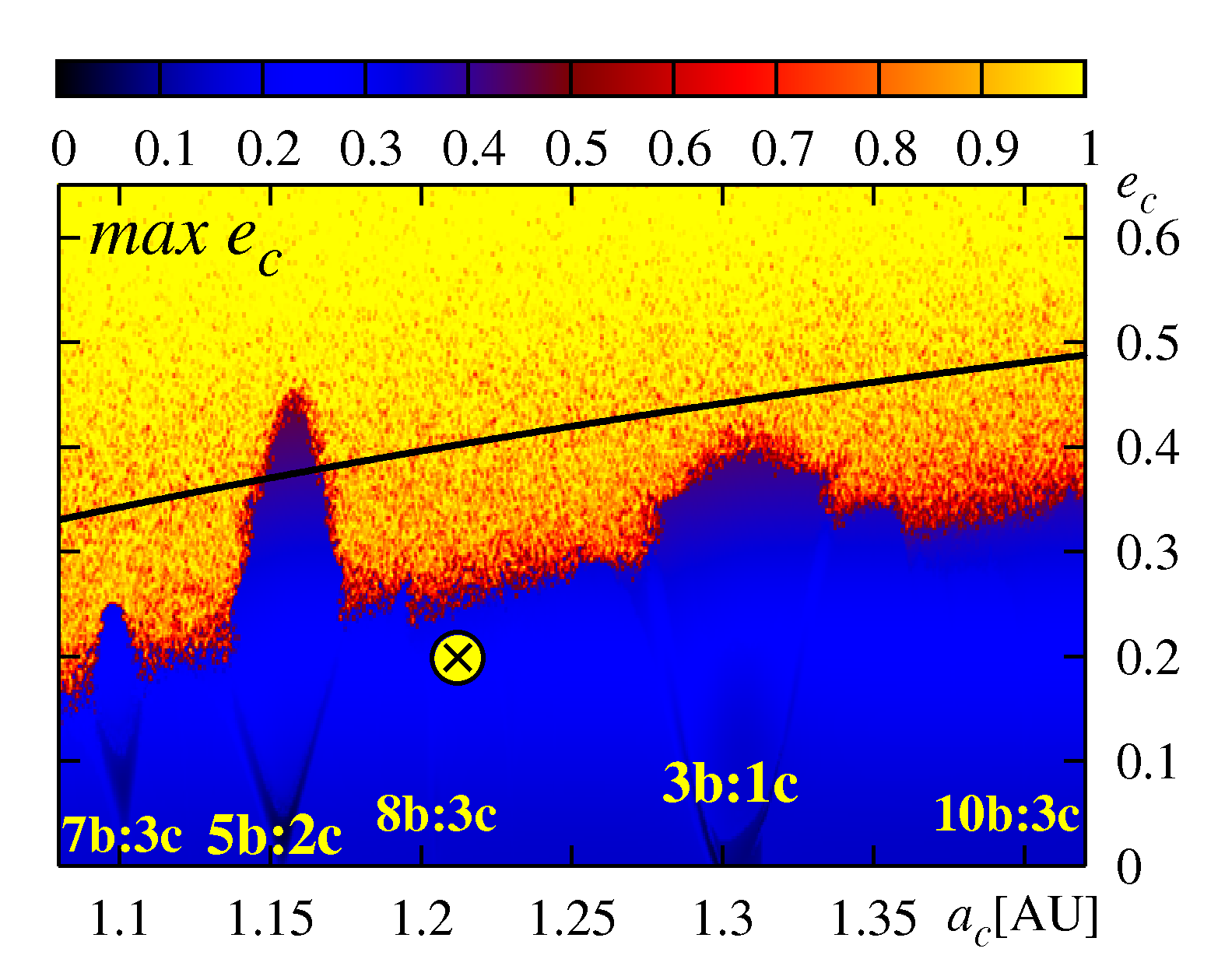}
}
}
\caption{\em 
Dynamical maps of putative \hdb{} coplanar configuration of two Jovian planets 
($\sim 0.5$--$0.9$~m$_{\idm{J}}$). The osculating elements  of the $N$-body
solution at the epoch of the first observation in  \cite{Cochran2007} are given
in Table~1 (fit~II) and marked with crossed circle. The top-left panel is for the Spectral Number. Colors mark
the stability regime: black is for regular solutions, yellow is for strongly
chaotic solutions. The top-right panel is for the maximal amplitude of apsidal
angle  $\Delta\varpi = \varpi_{\idm{c}}-\varpi_{\idm{b}}$. Panels in the bottom
row are for the $\max e$ indicator  (i.e., the maximal eccentricity attained
after the integration period $\sim 30,000$~yr). The most prominent
mean motion resonances between the planets are labeled.
}
\label{fig:fig2}
\end{figure}
\else
\begin{figure}
\centerline{
\hbox{
  \includegraphics[width=52mm]{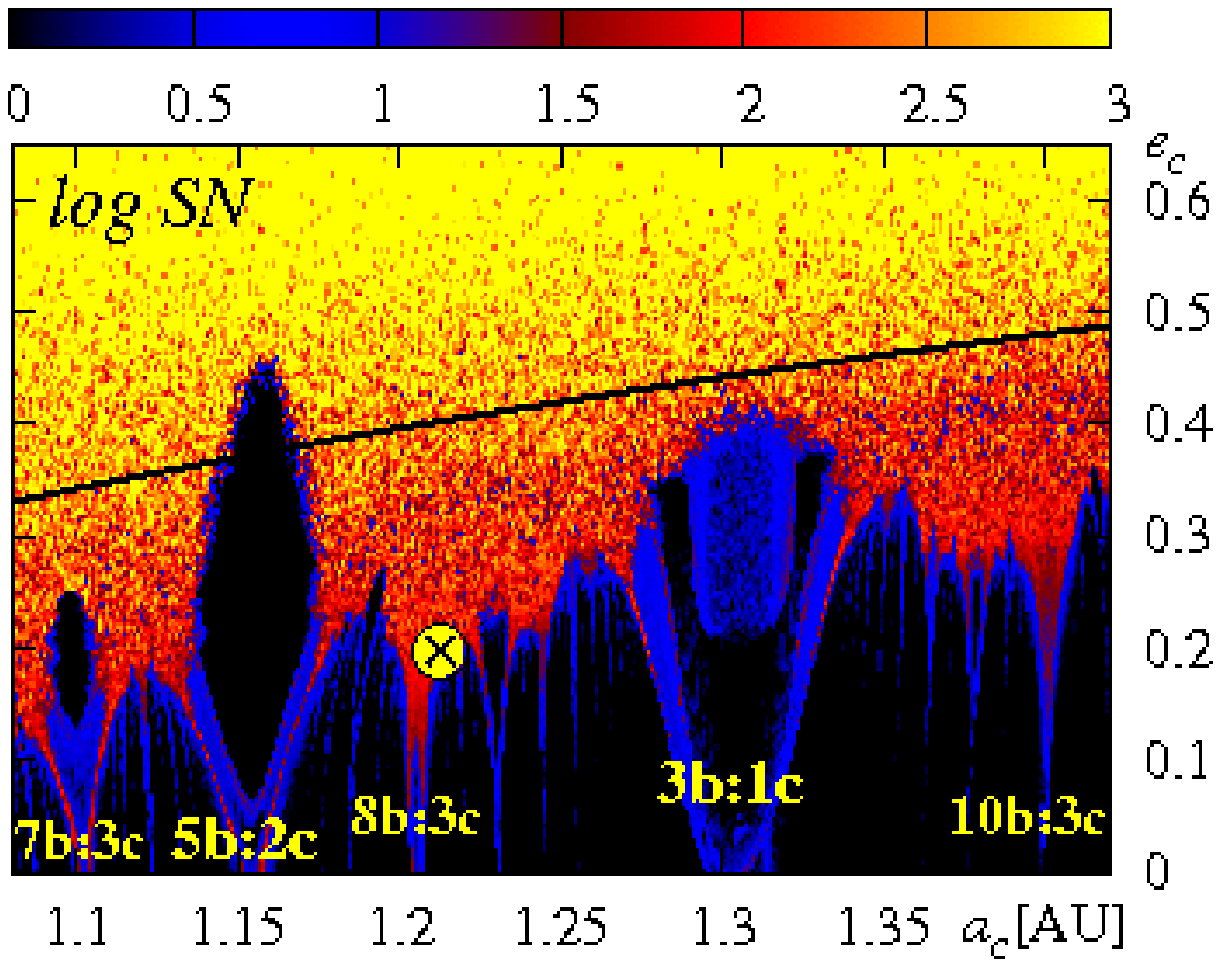}
  \hspace*{6mm}
  \includegraphics[width=52mm]{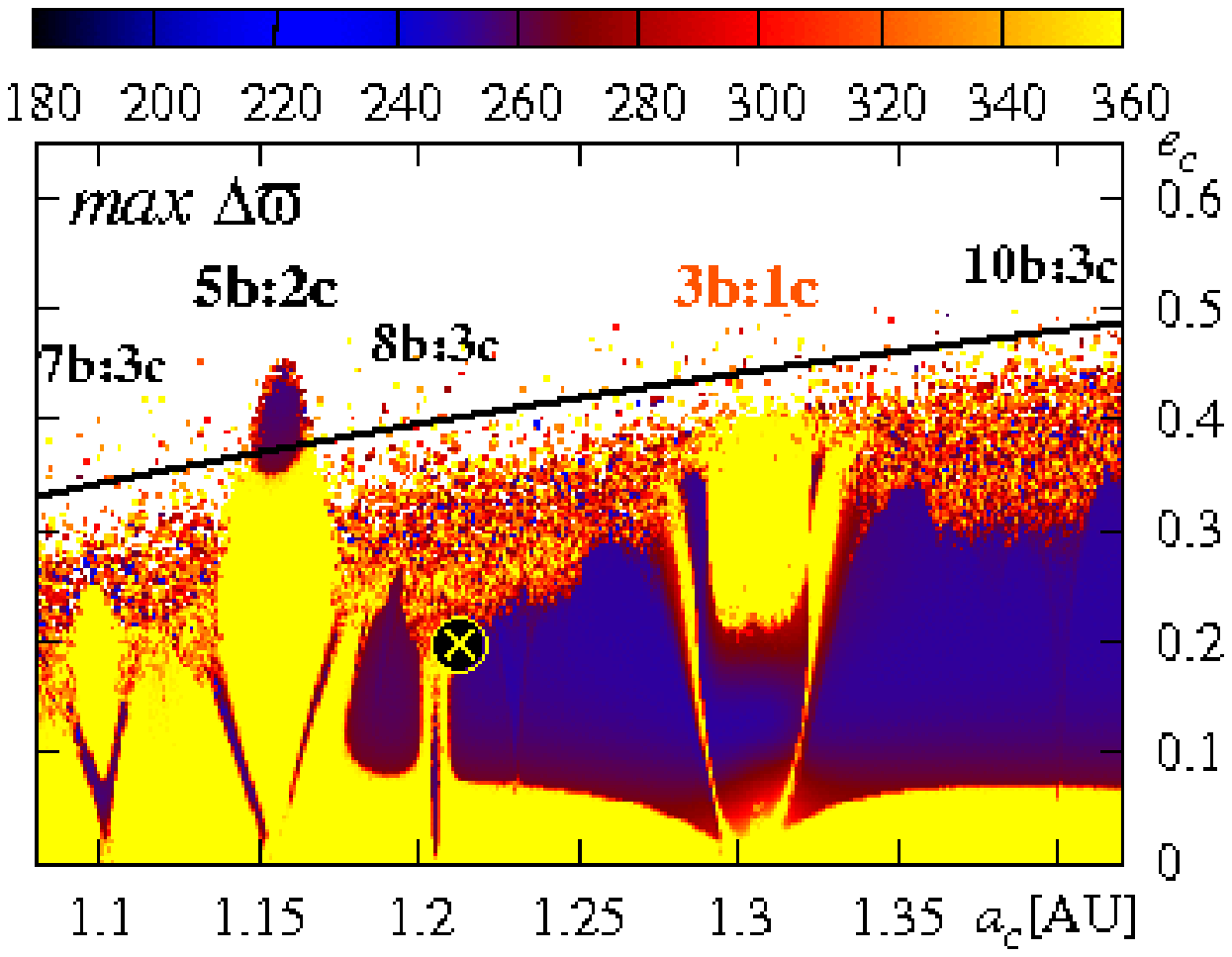}
}
}
\vspace*{0mm}
\centerline{
\hbox{
  \includegraphics[width=52mm]{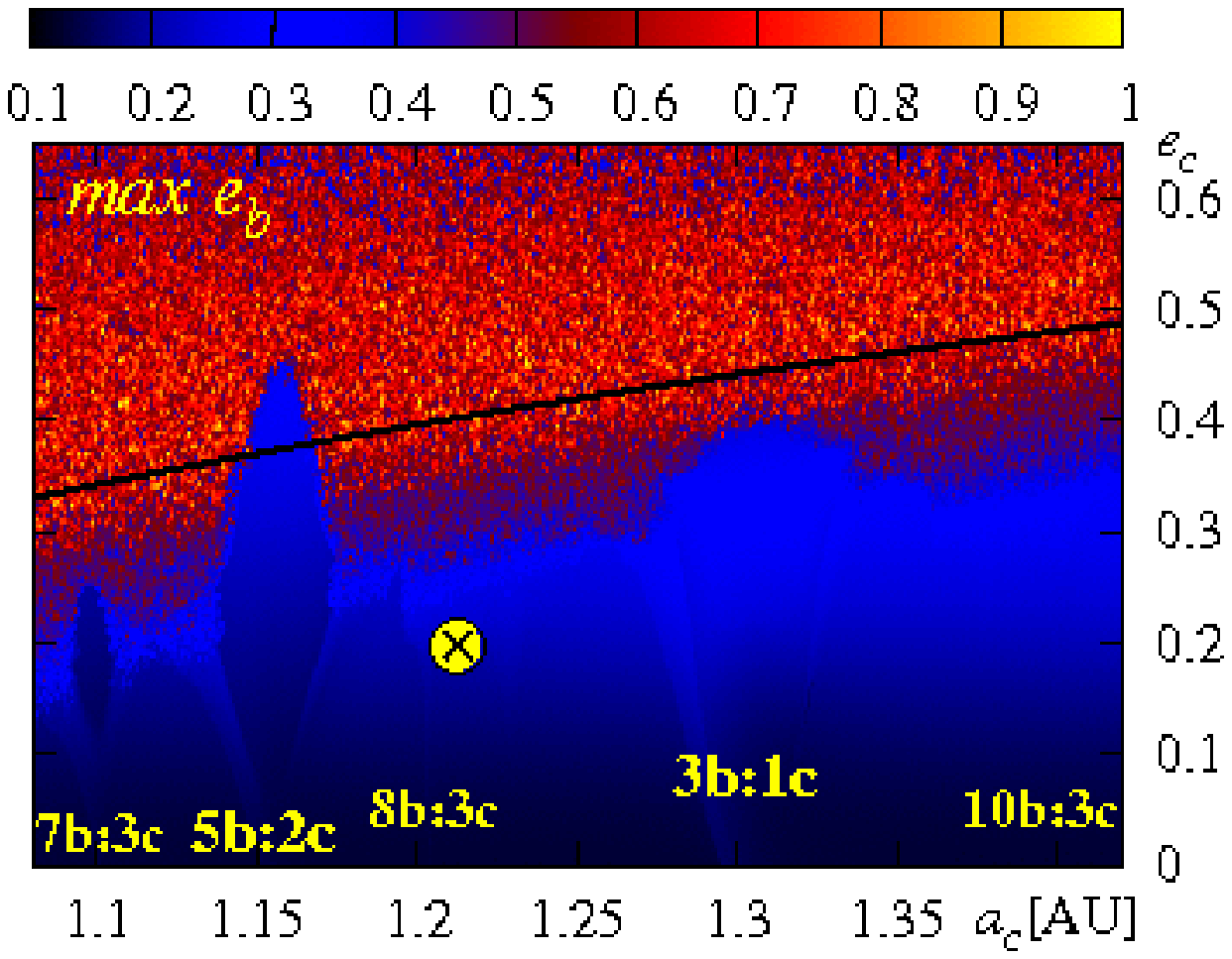}
   \hspace*{6mm}
  \includegraphics[width=52mm]{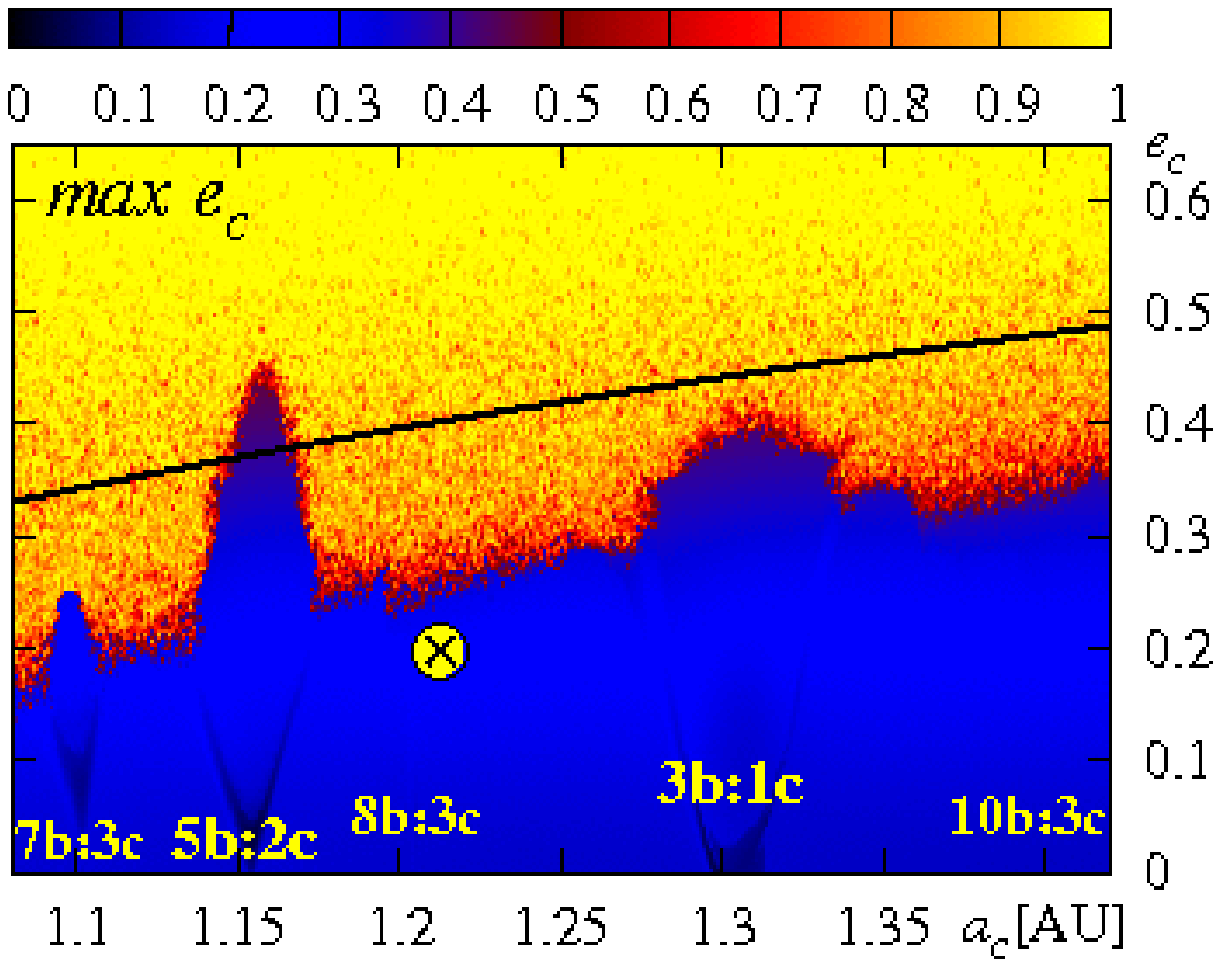}
}
}
\caption{\em 
Dynamical maps of putative \hdb{} coplanar configuration of two Jovian planets 
($\sim 0.5$--$0.9$~m$_{\idm{J}}$). The osculating elements  of the $N$-body
solution at the epoch of the first observation in  \cite{Cochran2007} are given
in Table~1 (fit~II) and marked with crossed circle. The top-left panel is for the Spectral Number. Colors mark
the stability regime: black is for regular solutions, yellow is for strongly
chaotic solutions. The top-right panel is for the maximal amplitude of apsidal
angle  $\Delta\varpi = \varpi_{\idm{c}}-\varpi_{\idm{b}}$. Panels in the bottom
row are for the $\max e$ indicator  (i.e., the maximal eccentricity attained
after the integration period $\sim 30,000$~yr). The most prominent
mean motion resonances between the planets are labeled.
}
\label{fig:fig2}
\end{figure}
\fi

To illustrate the hybrid optimization, we performed an independent search for
2-planet Keplerian fits assuming orbital periods in the range of [100,3000]~days
and eccentricities in the range of [0,0.8]. The ensemble of gathered fits is
shown in the top-left panel in Fig.\ref{fig:fig3}. To make the comparison
with the results of  systematic scanning more transparent, in this panel we also plot
contour levels of $1\sigma, 2\sigma$ and $3\sigma$ confidence intervals
seen in Fig.~\ref{fig:fig1}. Note that we plot only solutions in the range of
$P_{\idm{c}}\in [300,700]$~days (compare with the right panel in
Fig.~\ref{fig:fig1}).  The hybrid algorithm also reveals the minima seen in
Fig.~\ref{fig:fig1}. Besides, it detected one more minimum at $P_{\idm{c}}\sim
300$~days and moderate $e_{\idm{c}}\sim 0.1$, at the same depth.  This 
justifies the efficiency and robustness of the hybrid code. The algorithm not
only detects the best fits solutions but also helps to resolve to some extent
the shape of $\Chi$.

The best fits with $P_{\idm{c}}\sim 300$~days are both very unstable. It does
not necessarily mean that in their neighborhood some stable solutions do not
exist, so  it is the case in which the application of GAMP can be helpful.
Indeed, we can detect two clumps of stable fits (see the top-right panel of
Fig.~\ref{fig:fig3}) in the regime of large $e_{\idm{c}}$, both lying far over
the collision line. In one of these islands, we pick up a rigorously stable
solution with $\Chi \sim 1.14$ comparable to that one of the best fit~II
(Table~1) yielding  only marginally worse rms $\sim 6.3$~m/s. The evolution of
MEGNO and osculating elements in  this quasi-periodic configuration is shown in
Fig.~\ref{fig:fig4}. We notice extremely large variations of eccentricities up
to $0.8$. The system would be involved in 5:3~MMR protecting companions from
close encounter (see evolution of the critical argument of this resonance in the
bottom-right panel in Fig.~\ref{fig:fig4}).

Yet the fit parameters are determined within some error ranges that should be
interpreted with taking into account the structure of the phase space (see
Fig.~\ref{fig:fig2}). To illustrate this problem we examined more closely the
neighborhood of the best fit~II. At this time, we performed two experiments. In
the first search, we applied the hybrid code without stability constraints
driven by the ``usual'' $N$-body model of the RV. The results are illustrated in
the bottom-left panel in Fig.~\ref{fig:fig3}. The quality of fits within
$1\sigma,2\sigma$ and $3\sigma$ confidence intervals of the best fit (marked
with crossed circle; see Table~1, fit~II) is color coded with blue, light-blue
and gray, respectively. The best fits only marginally worse from the best one,
are marked in red. Curiously, the plot reveals a subtle structure with three
additional local minima of $\Chi$, in relatively small range of $a_{\idm{c}} \in
[1.1,1.3]$~AU. Moreover, these minima are spread over wide range of $e_{\idm{c}}
\in [0,0.7]$.  Simultaneously, this zone covers many low order resonances,
between 5:2 and 3:1~MMR and the $\Chi$ ``valley'' is crossed by the collision
line. Close to this line, the stability could be preserved only if the planets
are protected from close encounter through an MMR.

Now, we could examine the stability of every fit that we found but we choose  a new
search for stable solutions in a self-consistent manner with GAMP. The results are
shown in the bottom-right panel in Fig.~\ref{fig:fig3}. In this panel, we
overplot the stable solutions within $1\sigma,2\sigma$ and $3\sigma$  
confidence interval of the best stable fit~II over all solutions within
$3\sigma$ level which are found in the previous search (i.e., without stability
constraints). It now is evident that only a part of the $\Chi$ valley can
consist of dynamically stable solutions, nevertheless the  acceptable fits are
spread over significant range of $\Delta a_{\idm{c}} \sim 0.2$~AU. Apparently,
this error is quite small but  in fact it is large enough to cover a few
low-order MMRs. We conclude that the current set of RV data cannot fully
characterize the system state and new observations are required to constrain the
elements of the outer planet.

\ifpdf
\begin{figure}
\centerline{
\hbox{
  \includegraphics[width=60mm]{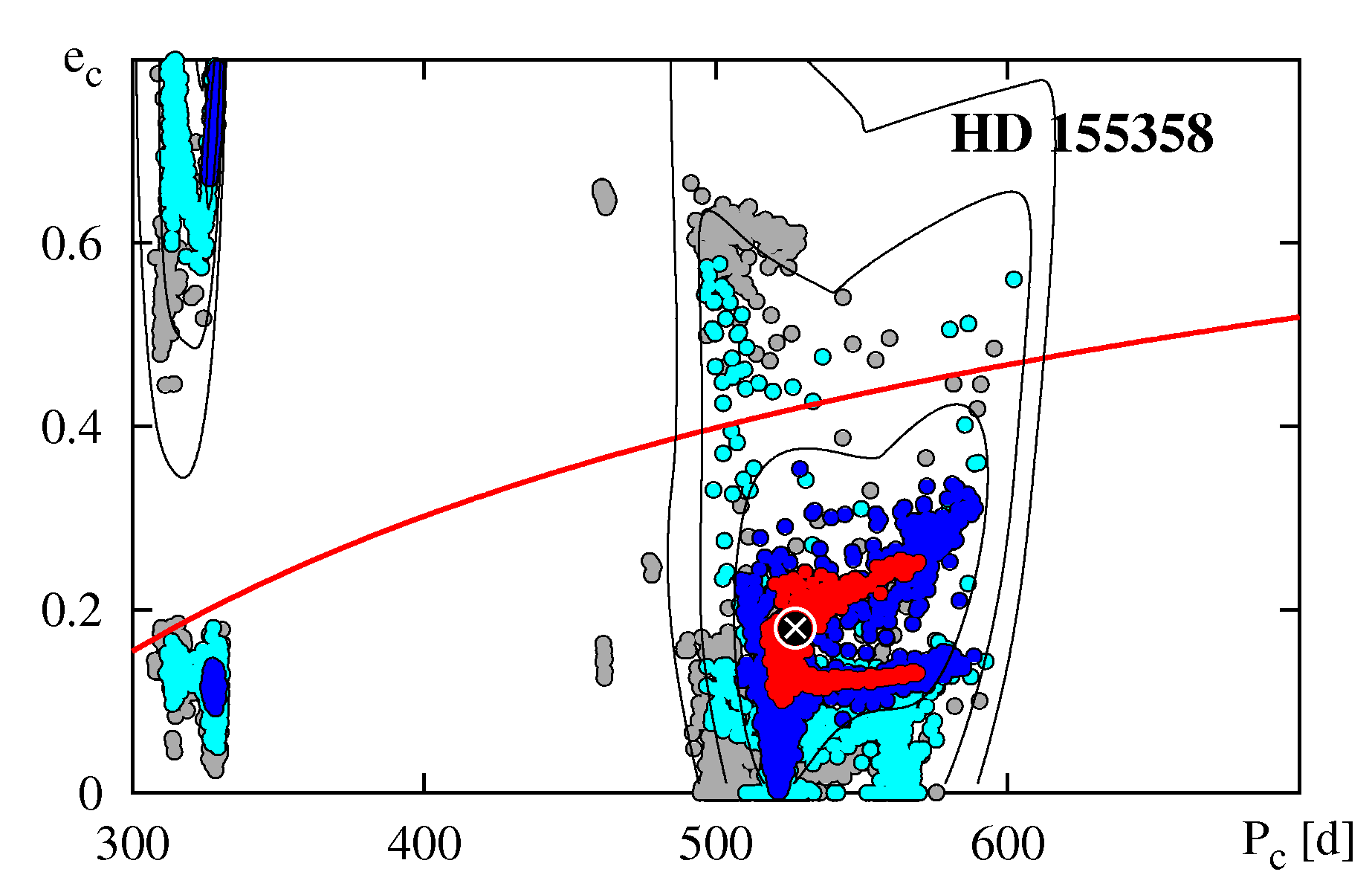}
  \hspace*{4mm}
  \includegraphics[width=60mm]{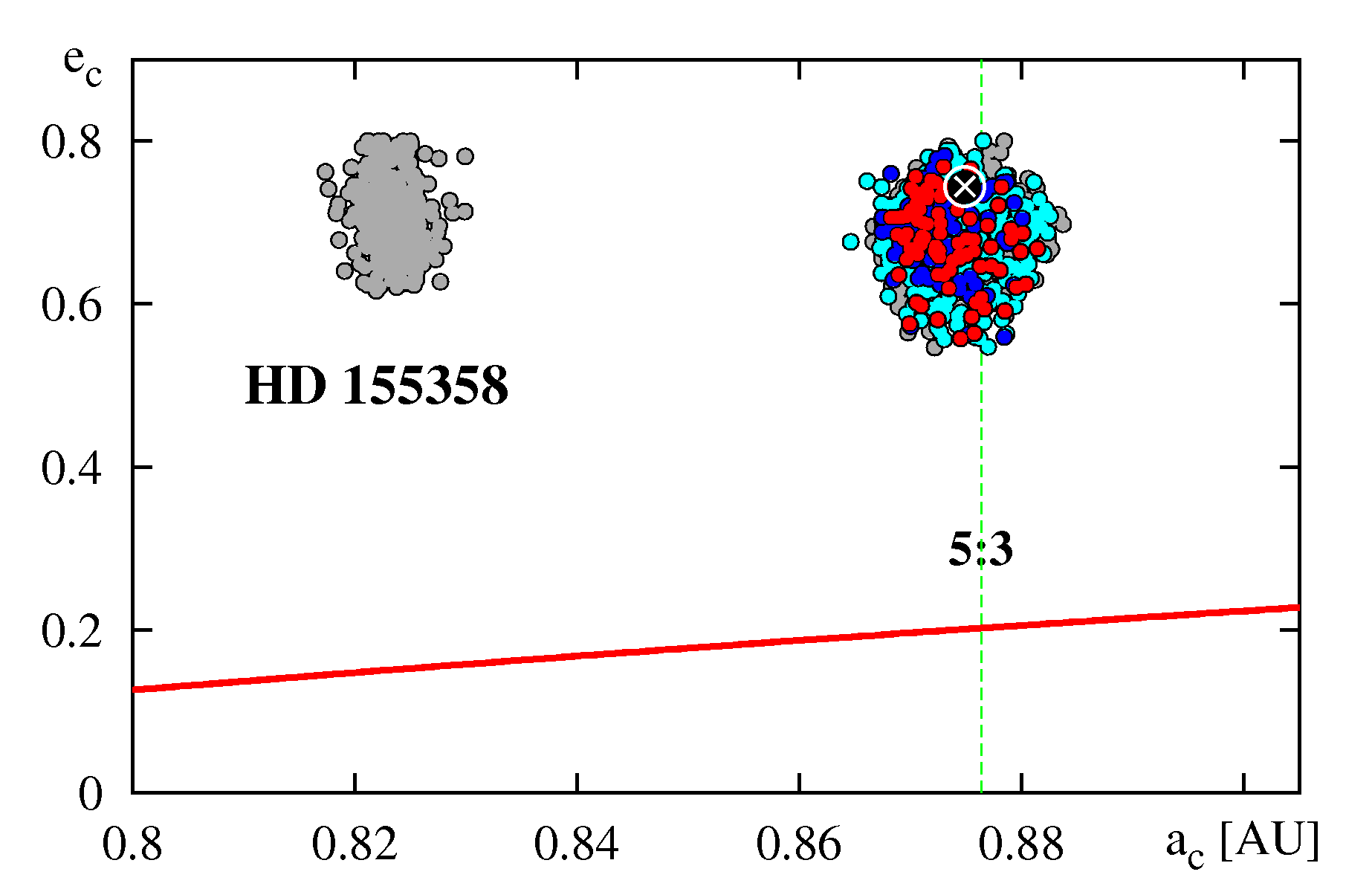}
}
}
\vspace*{0mm}
\centerline{
\hbox{
  \includegraphics[width=60mm]{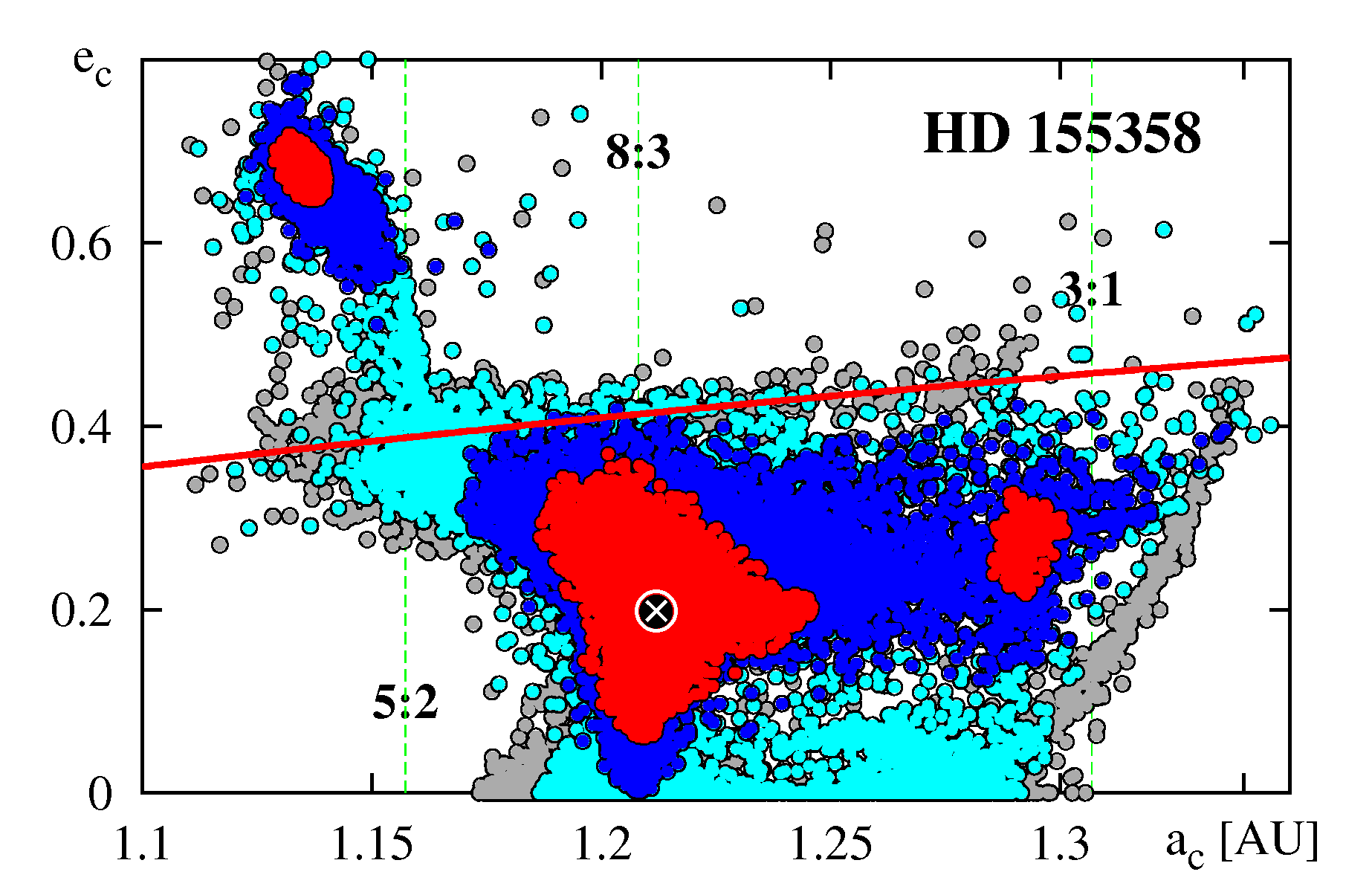}
  \hspace*{4mm}
  \includegraphics[width=60mm]{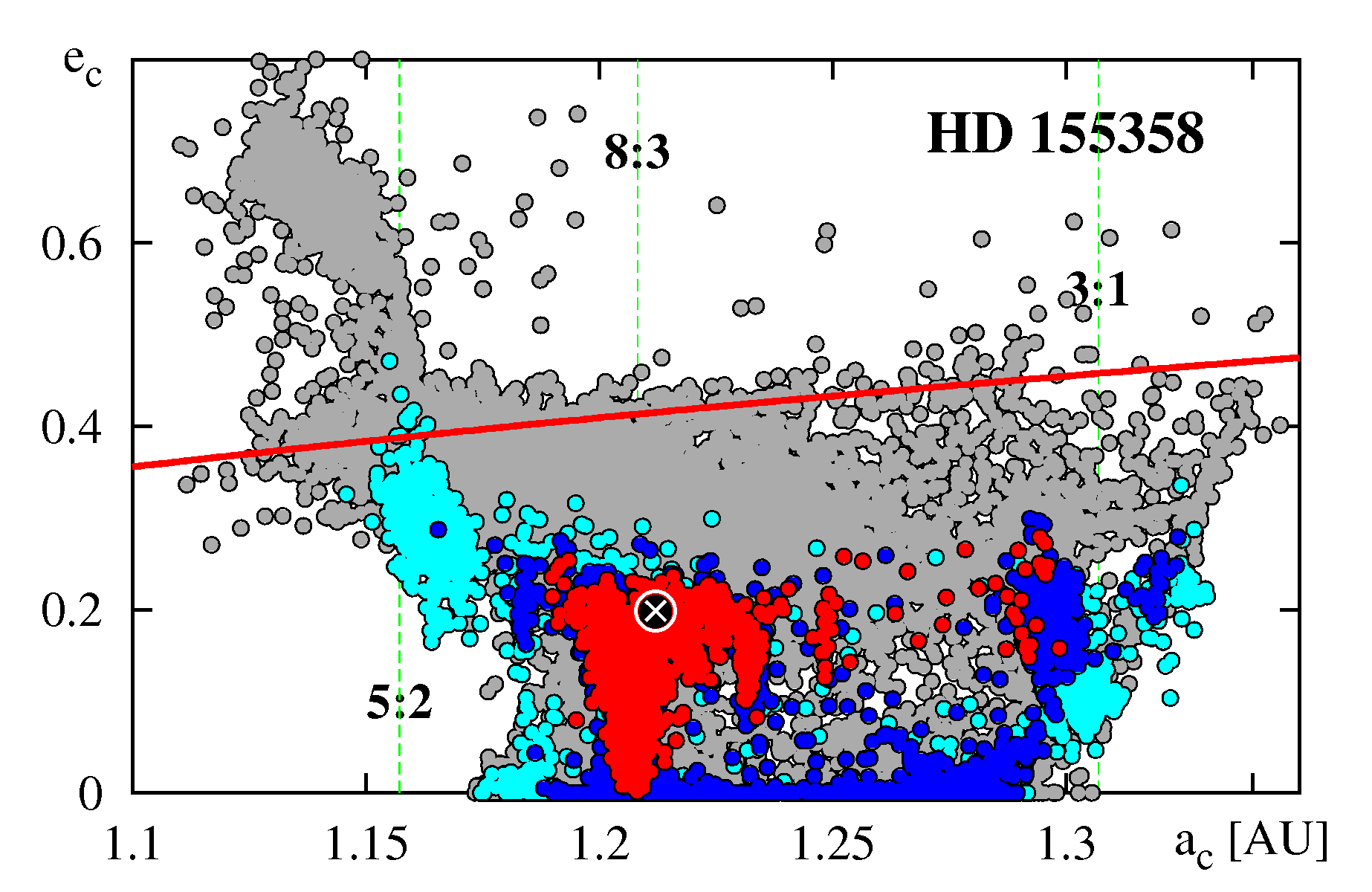}
}
}
\caption{\em 
The top-left panel is for the $\Chi$ of the 2-planet Keplerian solutions  to the
\hdb{} RV data published in Cochran et al. (2007)  derived with the  hybrid
algorithm. The fit parameters are projected  onto the
$(P_{\idm{c}},e_{\idm{c}})$-plane. Their quality is color coded: dark blue is
for $1\sigma$-, light-blue is for $2\sigma$- and grey is for
$3\sigma$-confidence levels of the best fit marked with crossed circle. 
Contours are for the confidence levels  obtained in the systematic scan (see the
right panel in Fig.~1). The top-right panel is for stable solutions in the range
$P_{\idm{c}} \sim 300$~days  derived with GAMP driven by the $N$-body model. 
The stable best-fit solution in this area is marked with crossed
circle, and its elements are given in Table~1, fit I. 
The bottom-left panel is for the ensemble of  fits gathered
with the hybrid algorithm driven by the $N$-body model of the RV data around the
dominant minimum of $\Chi$. The bottom-right panel illustrates stable solutions
gathered in the GAMP search. The set of solutions
within $3\sigma$ of the best stable fit (elements given in Table~1,
fit~II) but obtained without stability constraints,
as shown in the bottom-left panel, are again shown as a gray-filled contour.
Approximate positions of the most prominent MMRs are labeled.
}
\label{fig:fig3}
\end{figure}
\else
\begin{figure}
\centerline{
\hbox{
  \includegraphics[width=60mm]{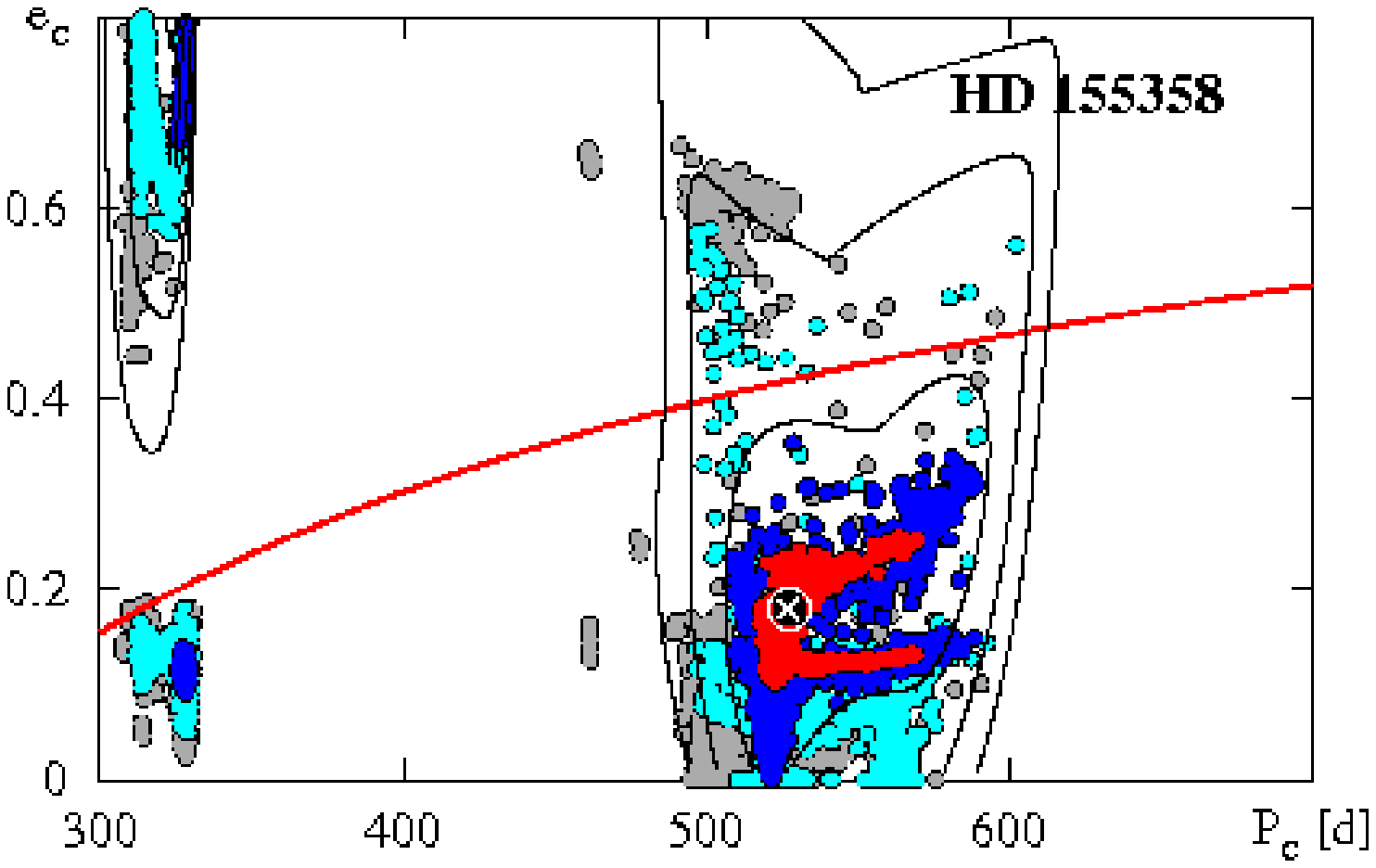}
  \hspace*{4mm}
  \includegraphics[width=60mm]{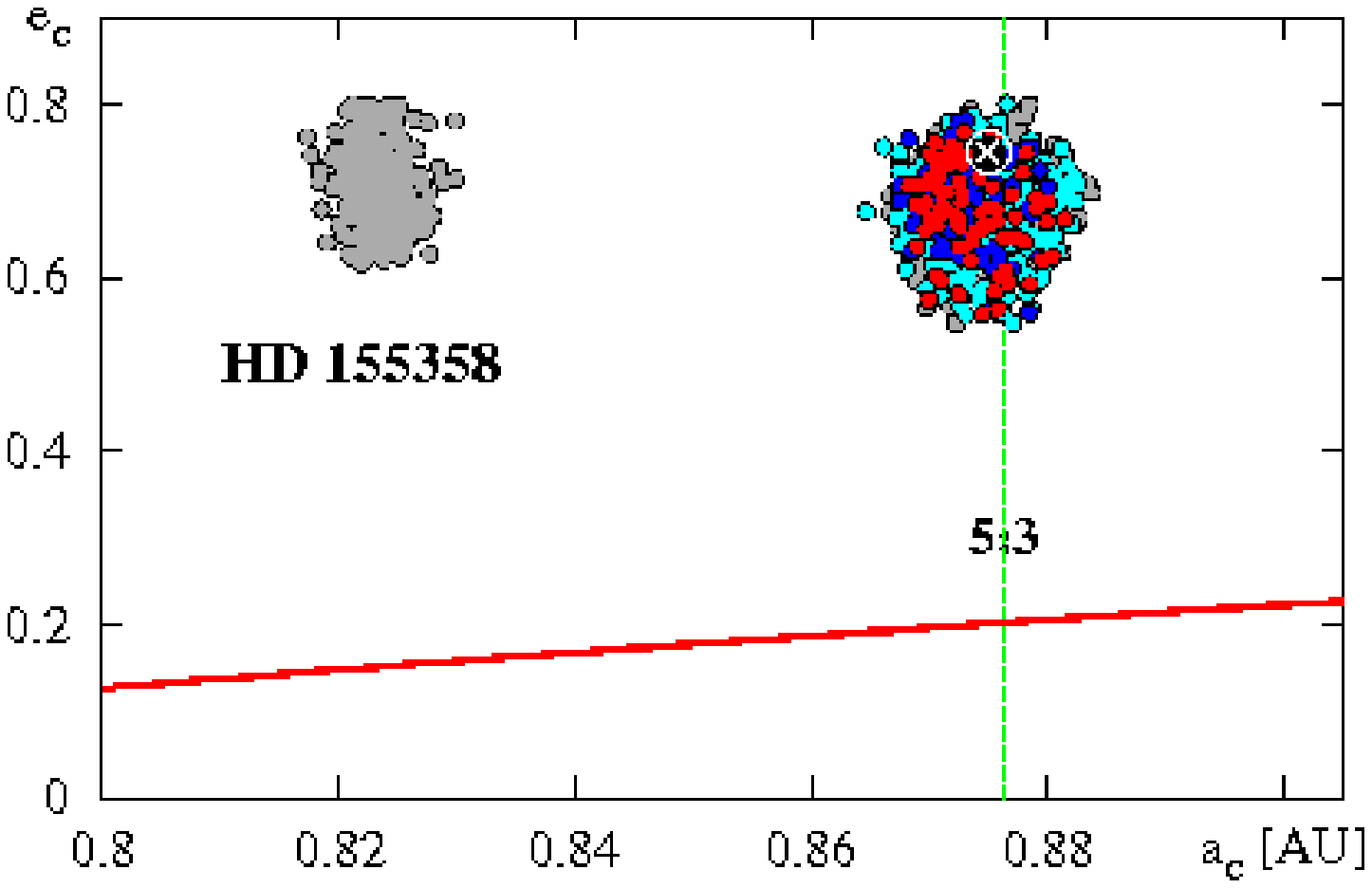}
}
}
\vspace*{0mm}
\centerline{
\hbox{
  \includegraphics[width=60mm]{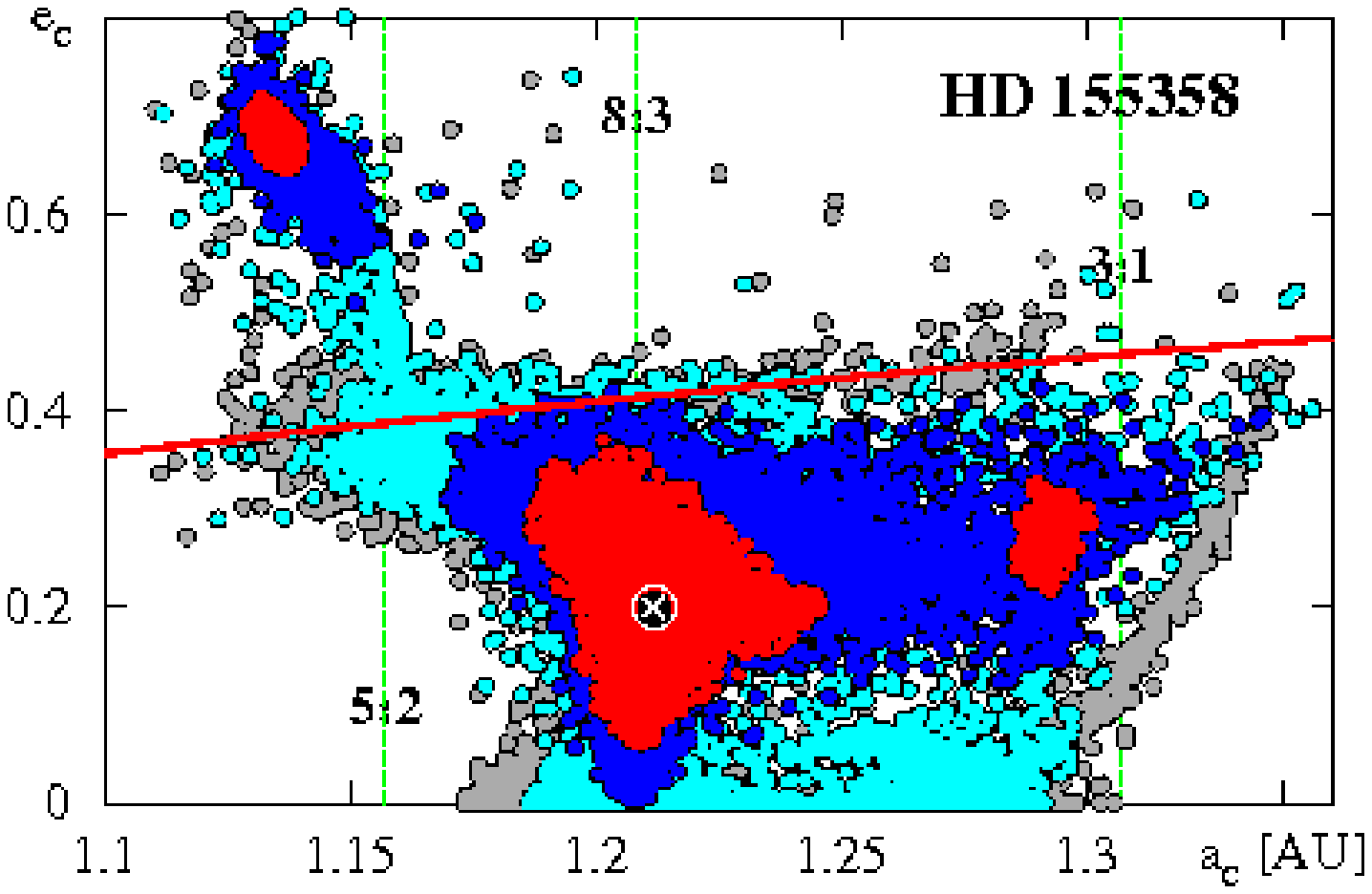}
  \hspace*{4mm}
  \includegraphics[width=60mm]{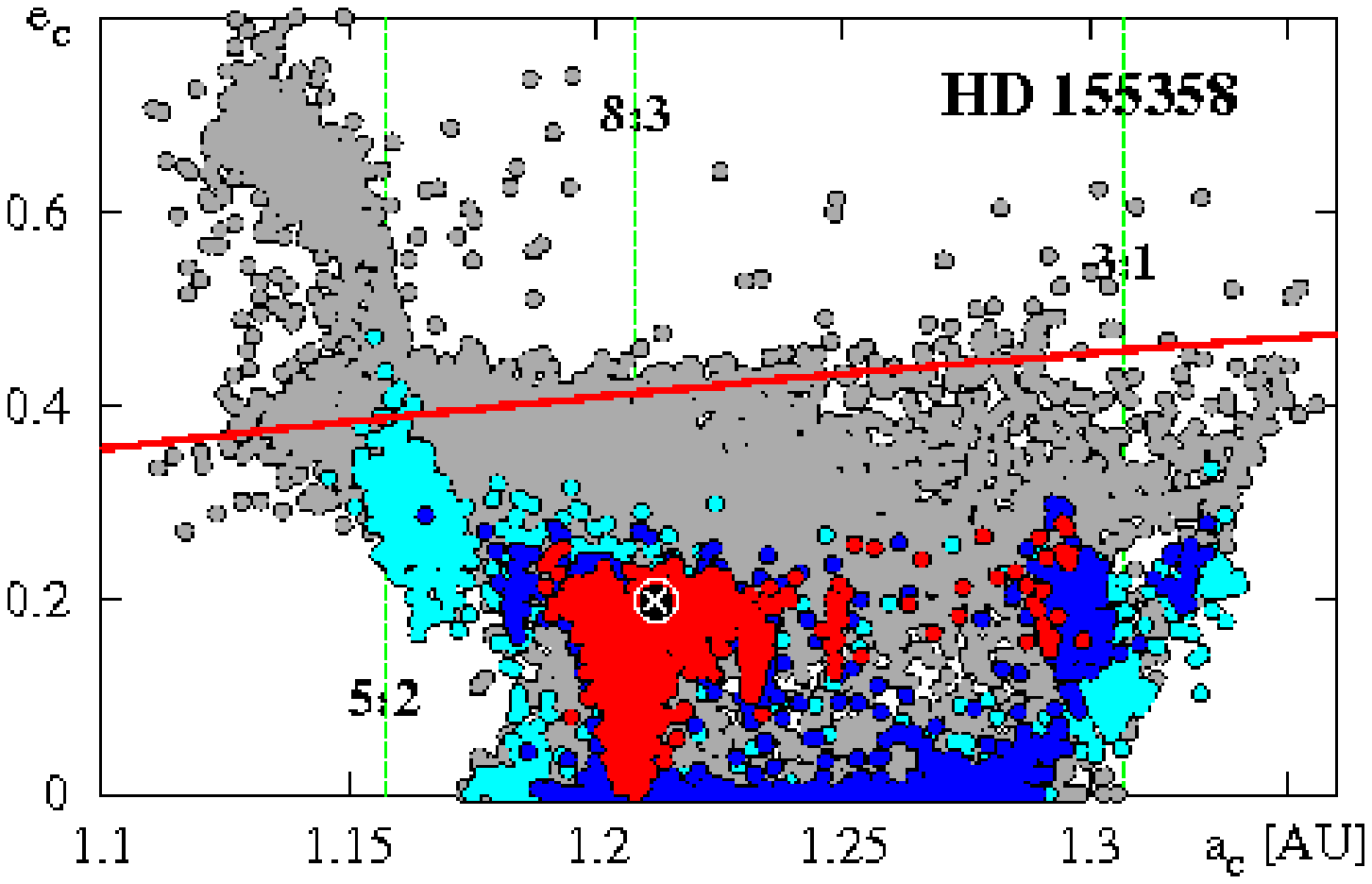}
}
}
\caption{\em 
The top-left panel is for the $\Chi$ of the 2-planet Keplerian solutions  to the
\hdb{} RV data published in Cochran et al. (2007)  derived with the  hybrid
algorithm. The fit parameters are projected  onto the
$(P_{\idm{c}},e_{\idm{c}})$-plane. Their quality is color coded: dark blue is
for $1\sigma$-, light-blue is for $2\sigma$- and grey is for
$3\sigma$-confidence levels of the best fit marked with crossed circle. 
Contours are for the confidence levels  obtained in the systematic scan (see the
right panel in Fig.~1). The top-right panel is for stable solutions in the range
$P_{\idm{c}} \sim 300$~days  derived with GAMP driven by the $N$-body model. 
The stable best-fit solution in this area is marked with crossed
circle, and its elements are given in Table~1, fit I. 
The bottom-left panel is for the ensemble of  fits gathered
with the hybrid algorithm driven by the $N$-body model of the RV data around the
dominant minimum of $\Chi$. The bottom-right panel illustrates stable solutions
gathered in the GAMP search. The set of solutions
within $3\sigma$ of the best stable fit (elements given in Table~1,
fit~II) but obtained without stability constraints,
as shown in the bottom-left panel, are again shown as a gray-filled contour.
Approximate positions of the most prominent MMRs are labeled.
}
\label{fig:fig3}
\end{figure}
\fi

Finally, both Keplerian and Newtonian 2-planet solutions lead to apparent excess
of the residuals, in particular at the end parts of the RV curve. It may
indicate that the 2-planet model does not fully explain the RV variability. In
particular, the  system may involve more than two planets. A heuristic argument
supporting such a claim may be the proximity of the best fits to the collision
line. We know similar cases, for instance  $\mu$~Arae
\citep{Gozdziewski2007a,Pepe2007}, or HD~37124~\citep{Vogt2005}. To check such
hypothesis we looked first for 3-planet Keplerian solutions with the hybrid
code. The best fit found  yields $\Chi \sim 0.87$ and significantly better rms
$\sim 4.6$~m/s but is unstable. The GAMP search yields stable configurations
with {\em quasi-circular} orbits of the outermost planets, yielding rms $\sim
5.5$~m/s. An example fit of this type, yielding $\Chi\sim 1.07$ and an rms $\sim
5.6$~m/s, given in terms of osculating element at the epoch of the first
observation in tuples of ($m$~[m$_{\idm{J}}$], $a$~[AU],$e$, $\omega$~[deg],
${\cal M}(t_0)$~[deg]) is the following  (0.115,   0.383,   0.025,   281.3,
154.4),  (0.770,   0.627,   0.040,   108.9, 173.7), (0.490,   1.187,   0.000,  
359.1,  65.7), for planets $d,b,c$ respectively, the offset $V_0=10.21$~m/s. We
note the small mass of the innermost planet. Its RV signal is at the level of
noise, so additional observations would be required to  confirm of withdrawn 
such a model.

\ifpdf
\begin{figure}
\centerline{
\hbox{
  \includegraphics[width=110mm]{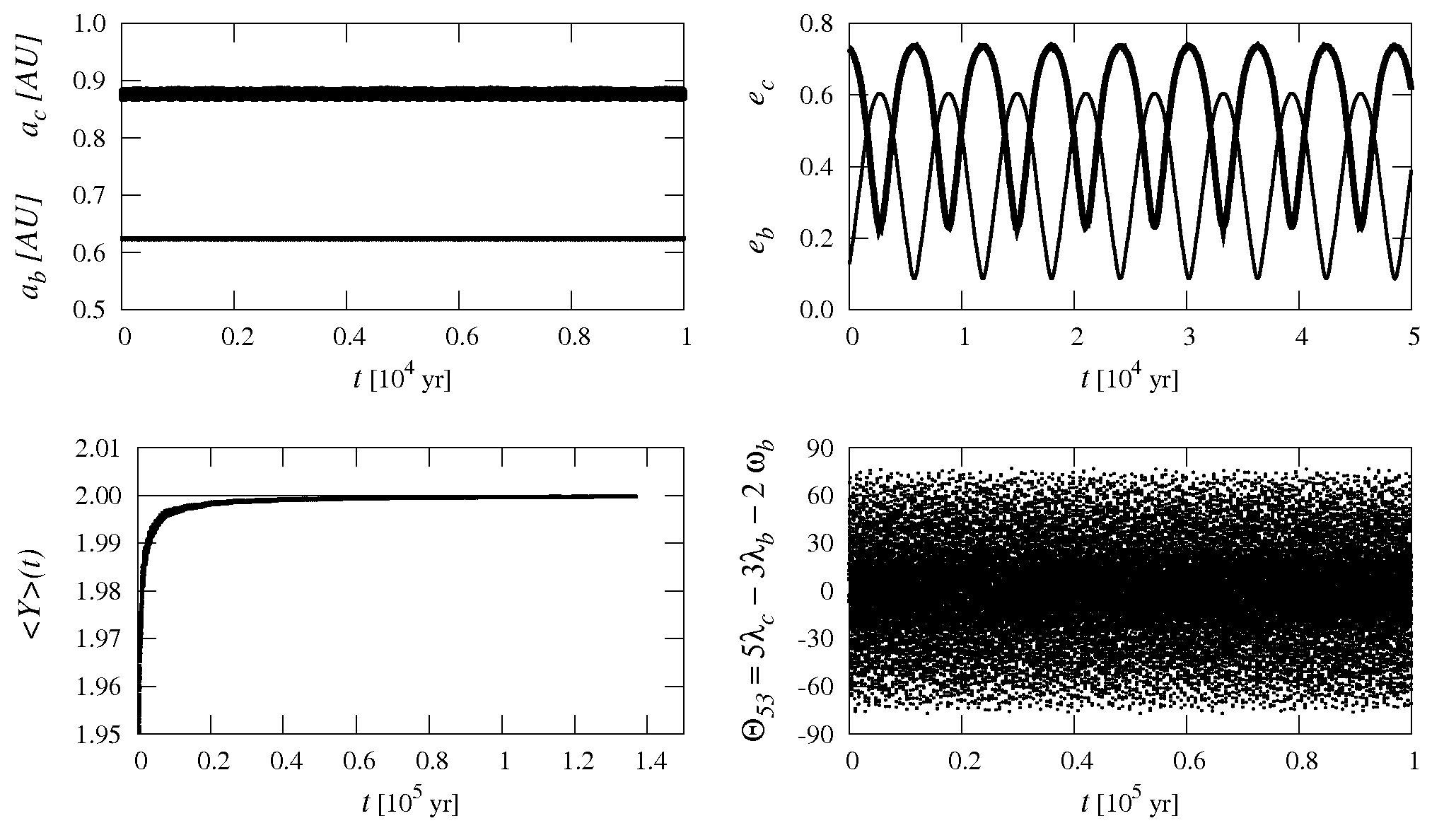}
}
}
\caption{\em 
The temporal evolution of  orbital elements in the 2-planet, Newtonian solution
to the RV data of \hdb{} related to highly eccentric orbits (see the top-left
panel in Fig.~\ref{fig:fig3}). The  osculating elements are given in Table~1
(fit~I). The top-left panel is for the semi-major axes, the top-right panel is
for the eccentricities. The MEGNO is plotted in the bottom-left panel. The
bottom-right panel is for the critical angle of the 5:3~MMR.
}
\label{fig:fig4}
\end{figure}
\else
\begin{figure}
\centerline{
\hbox{
  \includegraphics[width=110mm]{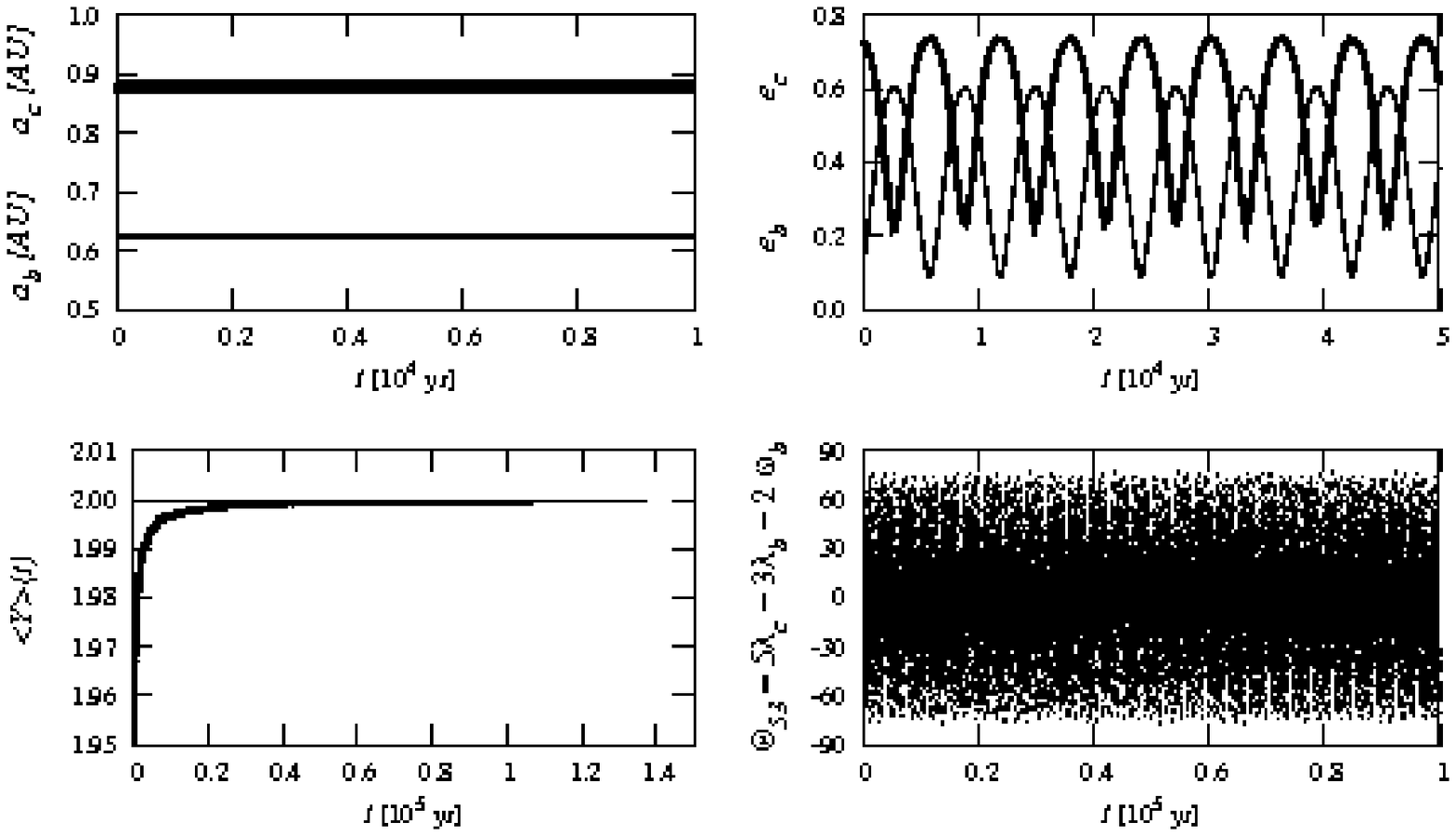}
}
}
\caption{\em 
The temporal evolution of  orbital elements in the 2-planet, Newtonian solution
to the RV data of \hdb{} related to highly eccentric orbits (see the top-left
panel in Fig.~\ref{fig:fig3}). The  osculating elements are given in Table~1
(fit~I). The top-left panel is for the semi-major axes, the top-right panel is
for the eccentricities. The MEGNO is plotted in the bottom-left panel. The
bottom-right panel is for the critical angle of the 5:3~MMR.
}
\label{fig:fig4}
\end{figure}
\fi

\section{Trojan planets in the  $\tau^1$~Gruis system?}
%
\cite{Laughlin2002a} predict that the reflex signal of a single planet in a
quasi-circular orbit may be also interpreted by two Jovian  Trojan planets,
i.e., two objects sharing similar orbits (involved in 1:1~MMR). That possibility
is intriguing  because stable Trojan companions to the  stars may be quite
common. It can be indicated by a number of stable Trojan configurations in the
Solar system. Some argue that they can be a frequent by-product of planet
formation and and/or dynamical evolution \citep{Laughlin2002a}.  However, the
genesis of Trojan planets  is not quite clear because on the contrary, there is
some evidence that formation of such  bodies could be difficult
\citep{Beauge2007}. Still, many authors expect that Trojan planets can exist
[see, for instance, the work of Dvorak et al. in this volume and references
therein, also \citep{Ford2006}]. 

Recently, we found a similar kind of ambiguity of the RV models concerning
2:1~MMR configurations. At present, we know five extrasolar systems presumably
involved in 2:1~MMR,  i.e., Gliese~876~\citep{Marcy2001},
HD~82943~\citep{Mayor2003}, HD~128311~\citep{Vogt2005},
\hda{}~\citep{Tinney2006}, and
$\mu$~Arae~\citep{Jones2002,Gozdziewski2007a,Pepe2007}. 
However, the 2:1~MMR model of the radial velocity observations can also be 
non-unique.  The periodogram  of the 2:1~MMR RV signal is very similar to that one
of the 1:1~MMR. Indeed, the RV variability of HD~128311 and HD~82943 can be
explained by highly inclined  systems in 1:1~MMR \citep{Gozdziewski2006x}. We
also  found that the RV of \hda{} can be  modeled  with two highly inclined
Jovian Trojans. The modeling of the 1:1~MMR  is a challenging problem because
Jupiter-like planets sharing eccentric orbits with similar semi-major axes
interact heavily and the collisional  configurations are generic. Hence, 
stability constraints are critical  in the search for optimal and  stable
configurations. This seems to be one of the best  applications of GAMP like
algorithms.

Among a few cases we analyzed so far, the $\tau^1$~Gruis  appears to be
a particularly interesting example of the possible  ``Jupiter on circular
orbit''--``two Trojans'' ambiguity. A Jovian companion to the G0 dwarf
$\tau^1$~Gruis  in a wide and almost circular orbit has been announced in the work
by \cite{Jones2002}. In our analysis, we use updated  RV data  comprising of
59~precision measurements \citep{Butler2006}. The best-fit single planet model
to these data yields $P_{\idm{b}}\sim1300$~days, and $e_{\idm{b}}\sim0.1$. 
We re-analyse the data to look for possible Trojan planet solutions. Curiously,
we quite easily found many {\em stable, coplanar} configurations involved in
1:1~MMR (Fig.~\ref{fig:fig5}) yielding similar or slightly better fit quality
(rms $\in [5,6]$~m/s).  The reflex signal of the 1:1~MMR
(Fig.~\ref{fig:fig6}) can hardly be distinguished from that of a
single-planet system.  The osculating elements  of the Trojans are given in
Table~1 (fit III).  In this case, both planets would move on quasi-circular 
orbits and these would be
{\em coplanar}. The dynamical maps shown in Fig.~\ref{fig:fig5} (also 
accompanying other best-fits solutions with acceptable quality which we found in
the search) illustrate the extreme variability of the 1:1~MMR islands.  The map
for the best-fit with elements in Table~1 is shown in the top-left  panel
of Fig.~\ref{fig:fig5}.  Another peculiar solution is illustrated in the top-right
map in Fig.~\ref{fig:fig5}. The initial eccentricities are moderate, and the
system would be found in extremely large island of stable motions. It spans
whole range of $e_{\idm{b}}$. The possibility of existence of such extended
stable  zones may strength the hypothesis of stable extrasolar Trojans.

\ifpdf
\begin{figure}
\centerline{
\hbox{
  \includegraphics[width=52mm]{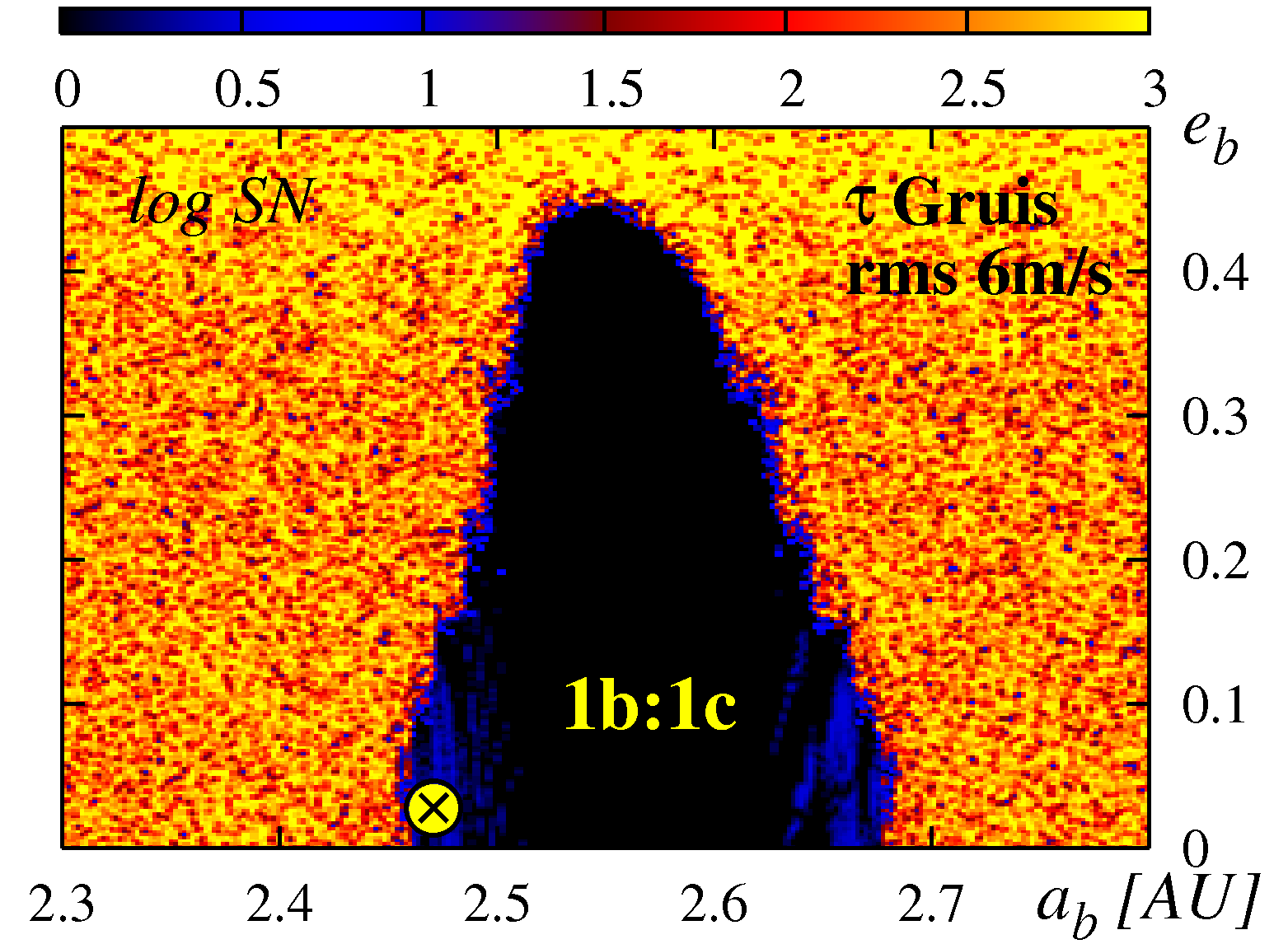}
  \hspace*{6mm}
  \includegraphics[width=52mm]{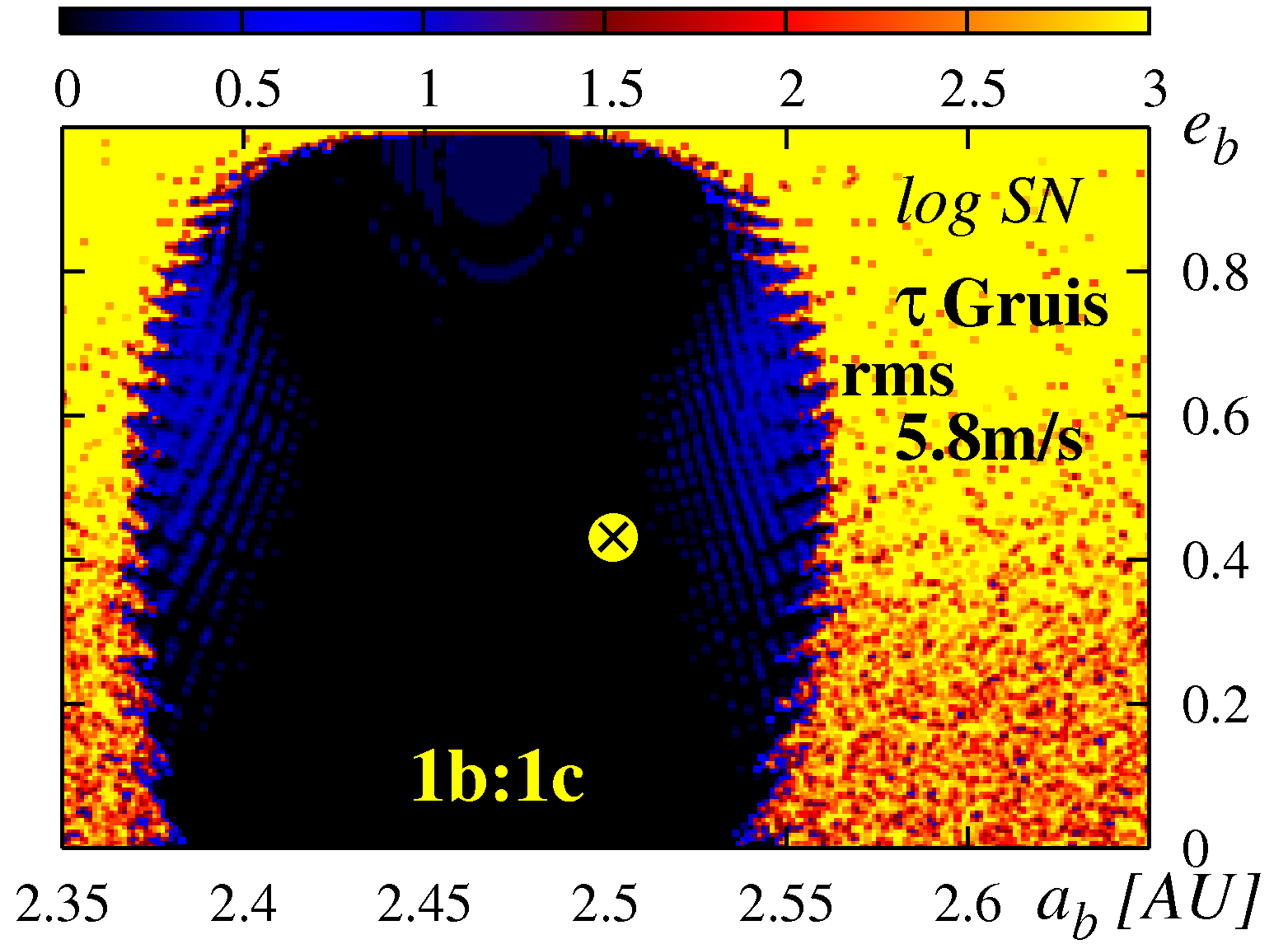}
}
}
\vspace*{0mm}
\centerline{
\hbox{
  \includegraphics[width=52mm]{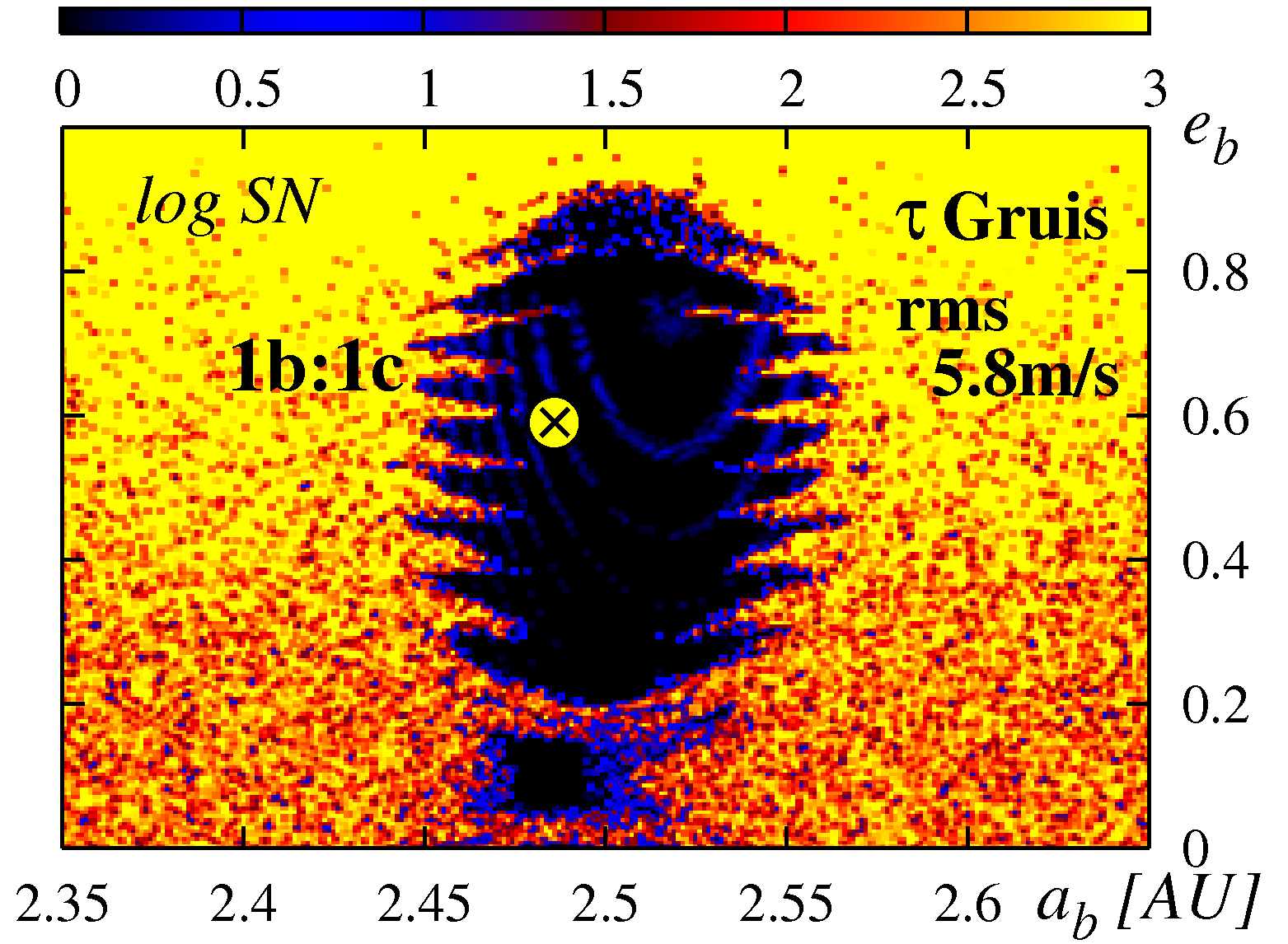}
   \hspace*{6mm}
  \includegraphics[width=52mm]{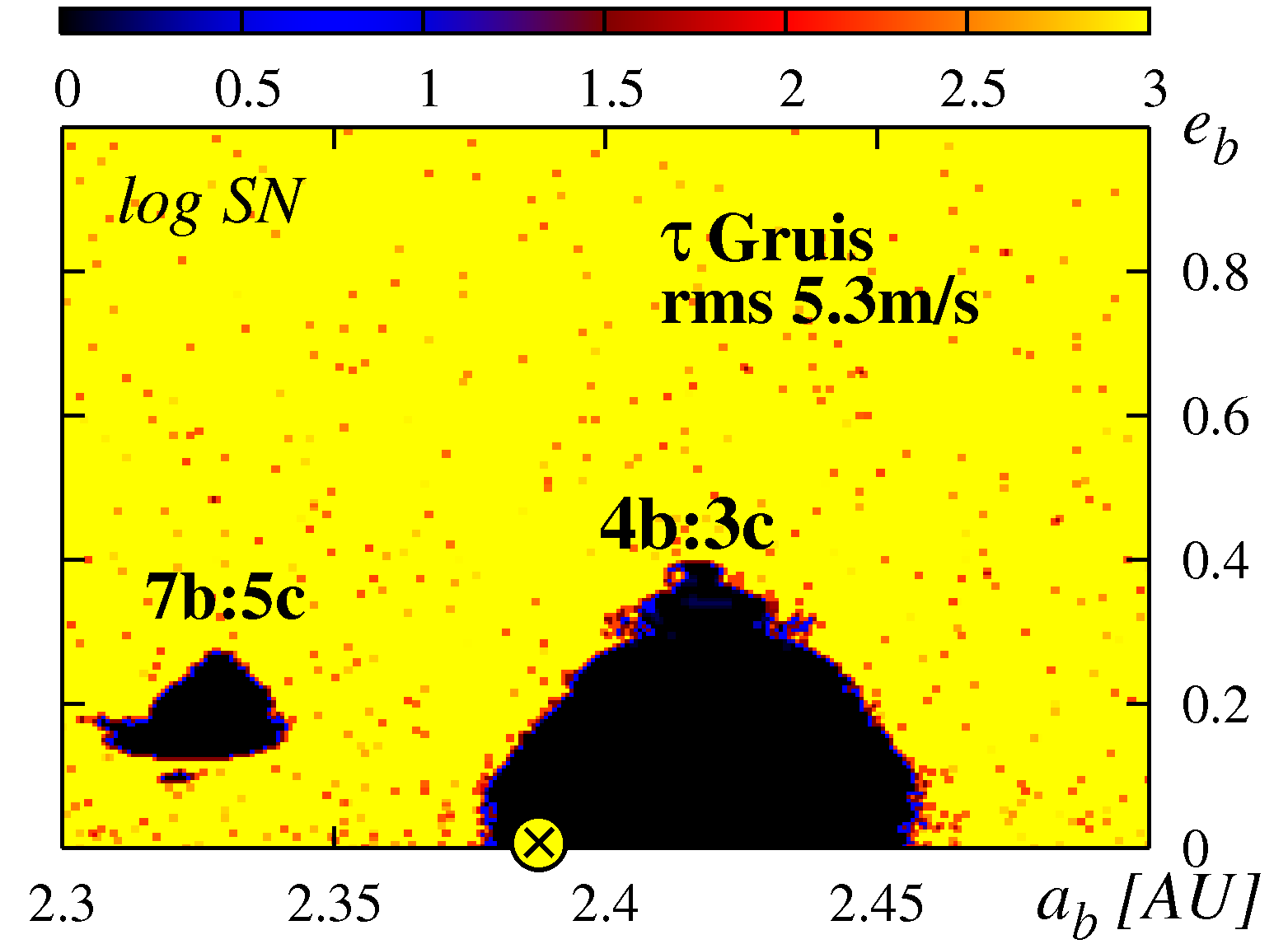}
}
}
\caption{\em
Dynamical maps of putative $\tau^1$~Gruis {coplanar, edge-on} configuration of two
Jovian planets ($\sim0.5$--$0.9$~m$_{\idm{J}}$) involved in low-order
resonances. The best-fits yield an rms $\sim5$--$6$~m/s and $\Chi\sim1$. Their
quality is similar to that of the single-planet solution (an rms about of 6~m/s).
}
\label{fig:fig5}
\end{figure}
\else
\begin{figure}
\centerline{
\hbox{
  \includegraphics[width=52mm]{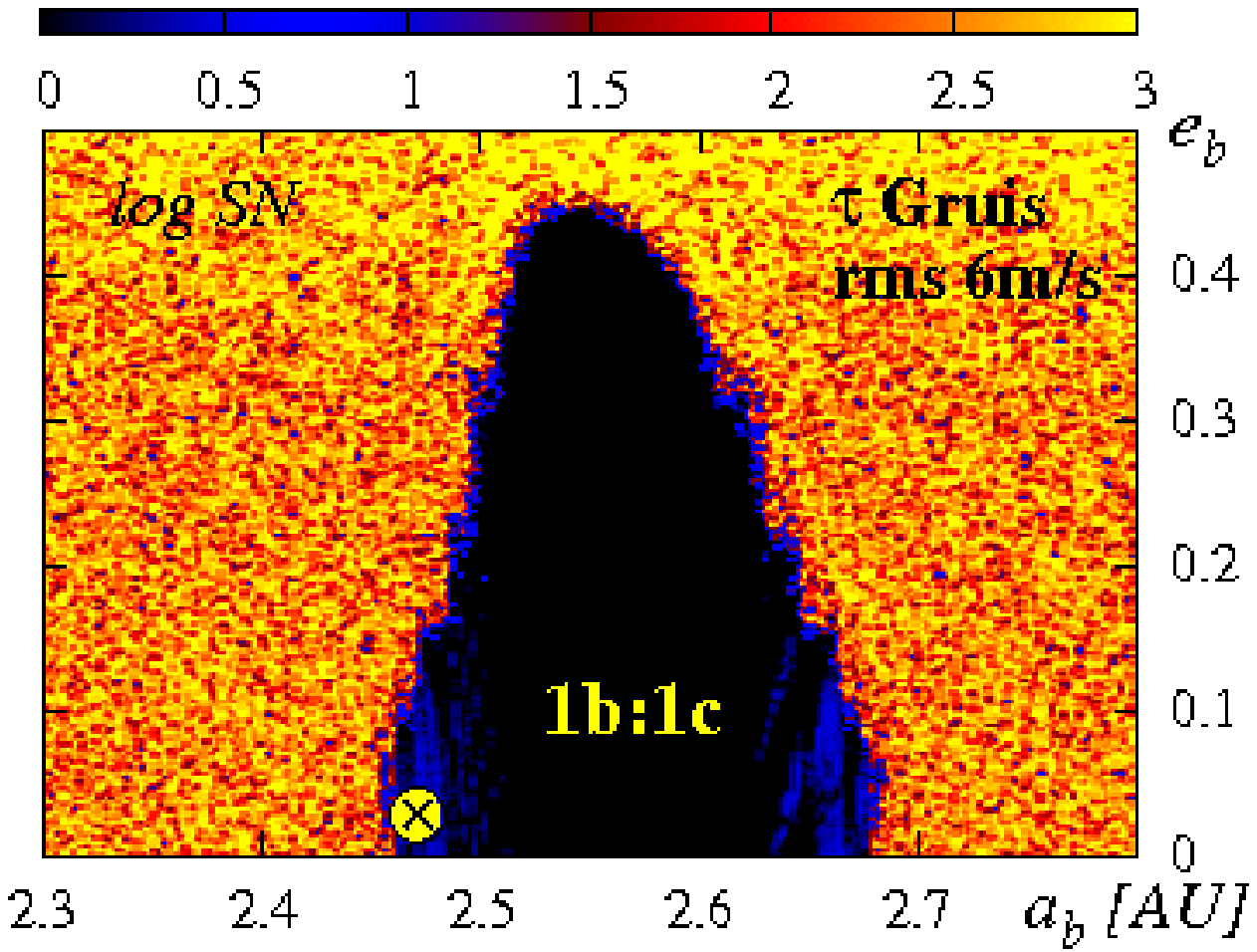}
  \hspace*{6mm}
  \includegraphics[width=52mm]{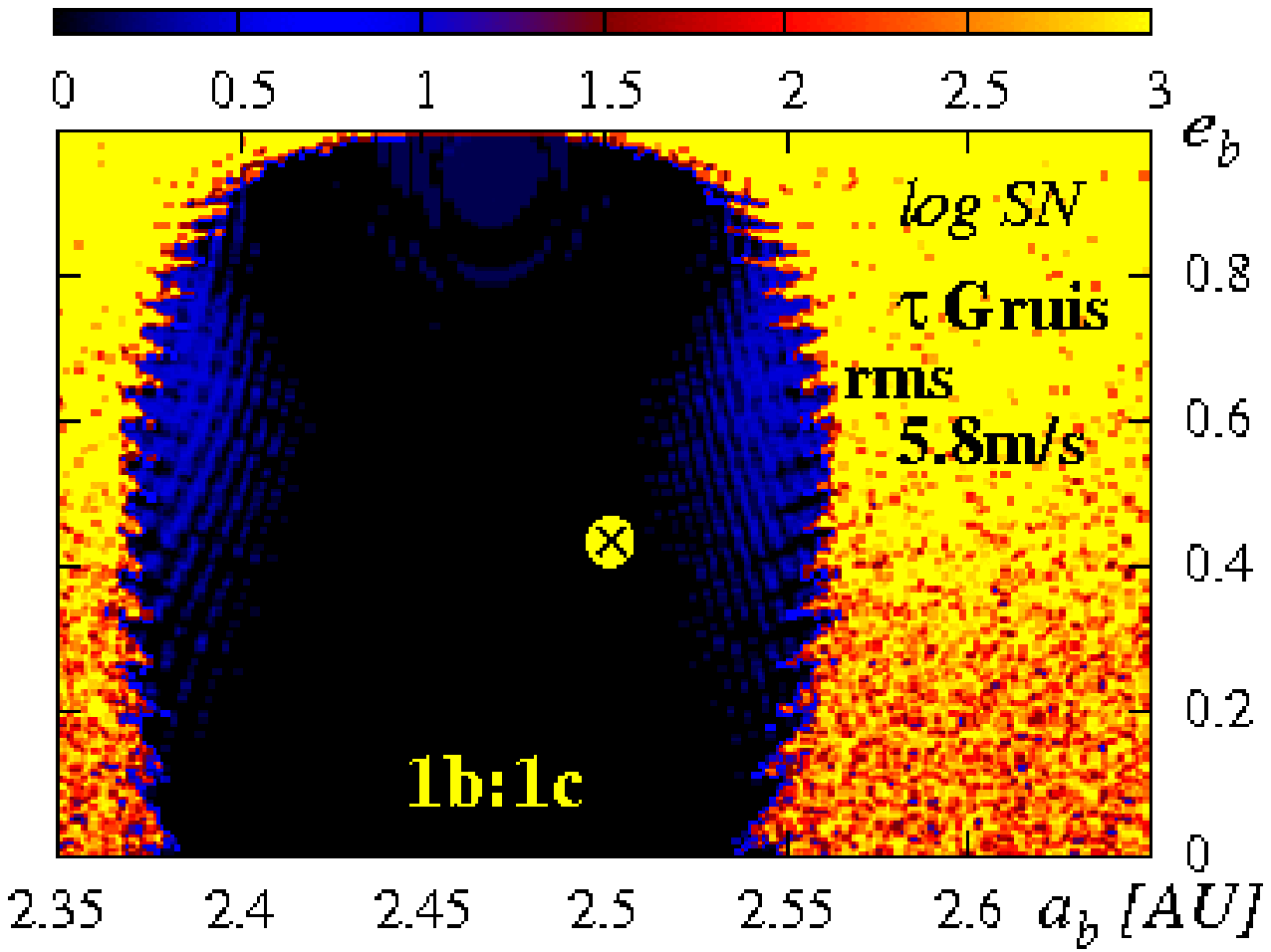}
}
}
\vspace*{0mm}
\centerline{
\hbox{
  \includegraphics[width=52mm]{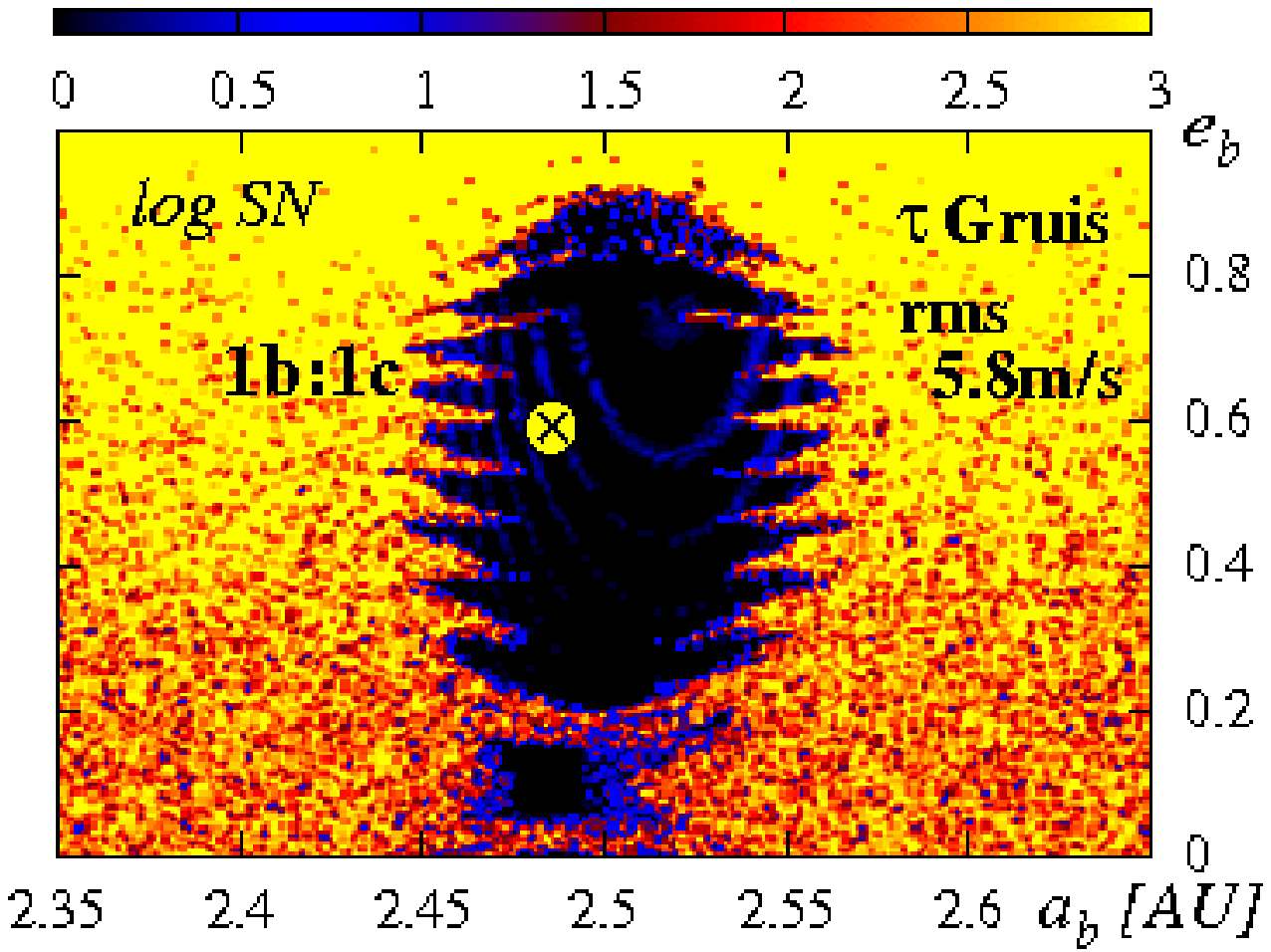}
   \hspace*{6mm}
  \includegraphics[width=52mm]{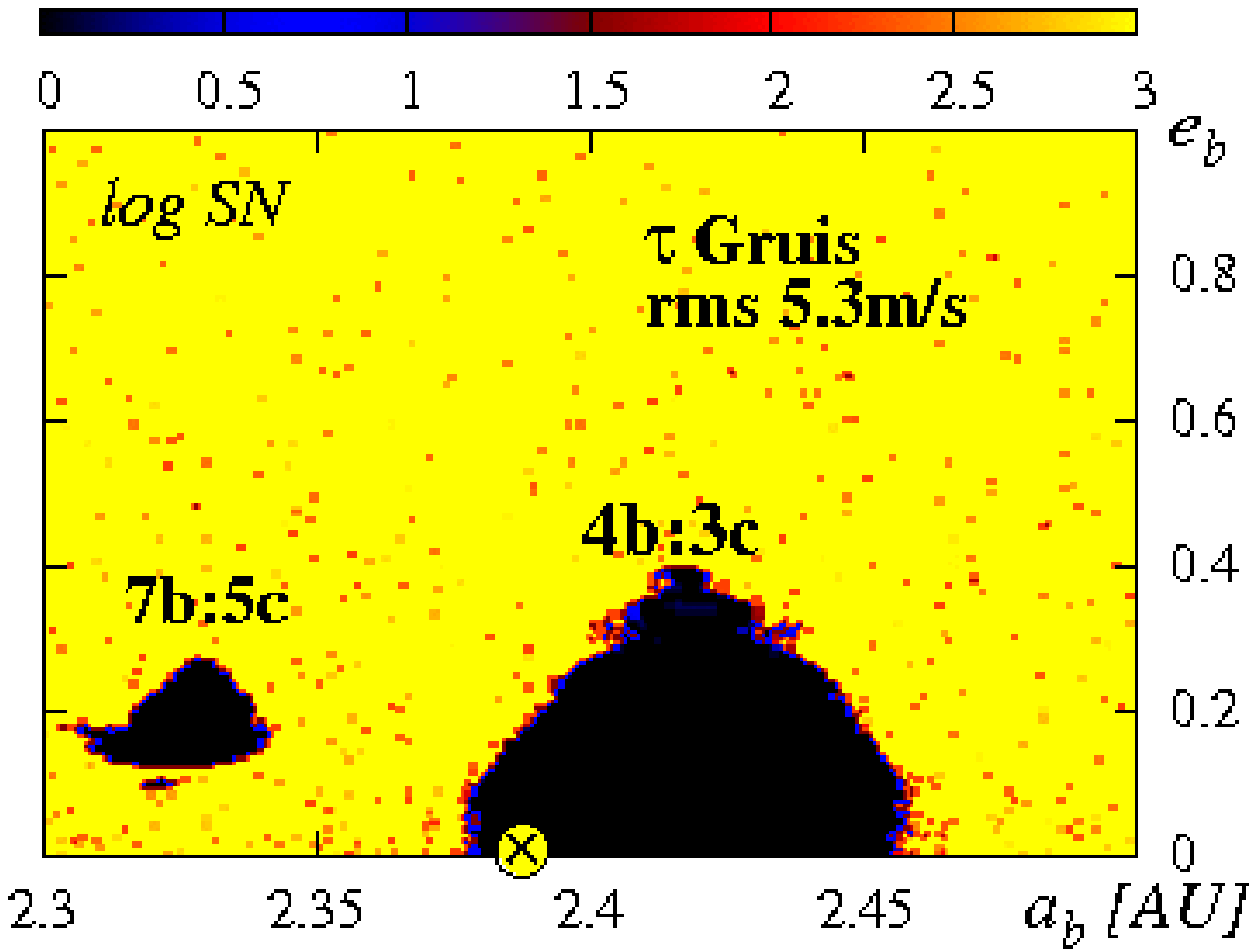}
}
}
\caption{\em
Dynamical maps of putative $\tau^1$~Gruis {coplanar, edge-on} configuration of two
Jovian planets ($\sim0.5$--$0.9$~m$_{\idm{J}}$) involved in low-order
resonances. The best-fits yield an rms $\sim5$--$6$~m/s and $\Chi\sim1$. Their
quality is similar to that of the single-planet solution (an rms about of 6~m/s).
}
\label{fig:fig5}
\end{figure}
\fi

\ifpdf
\begin{figure}
\centerline{\includegraphics[width=80mm]{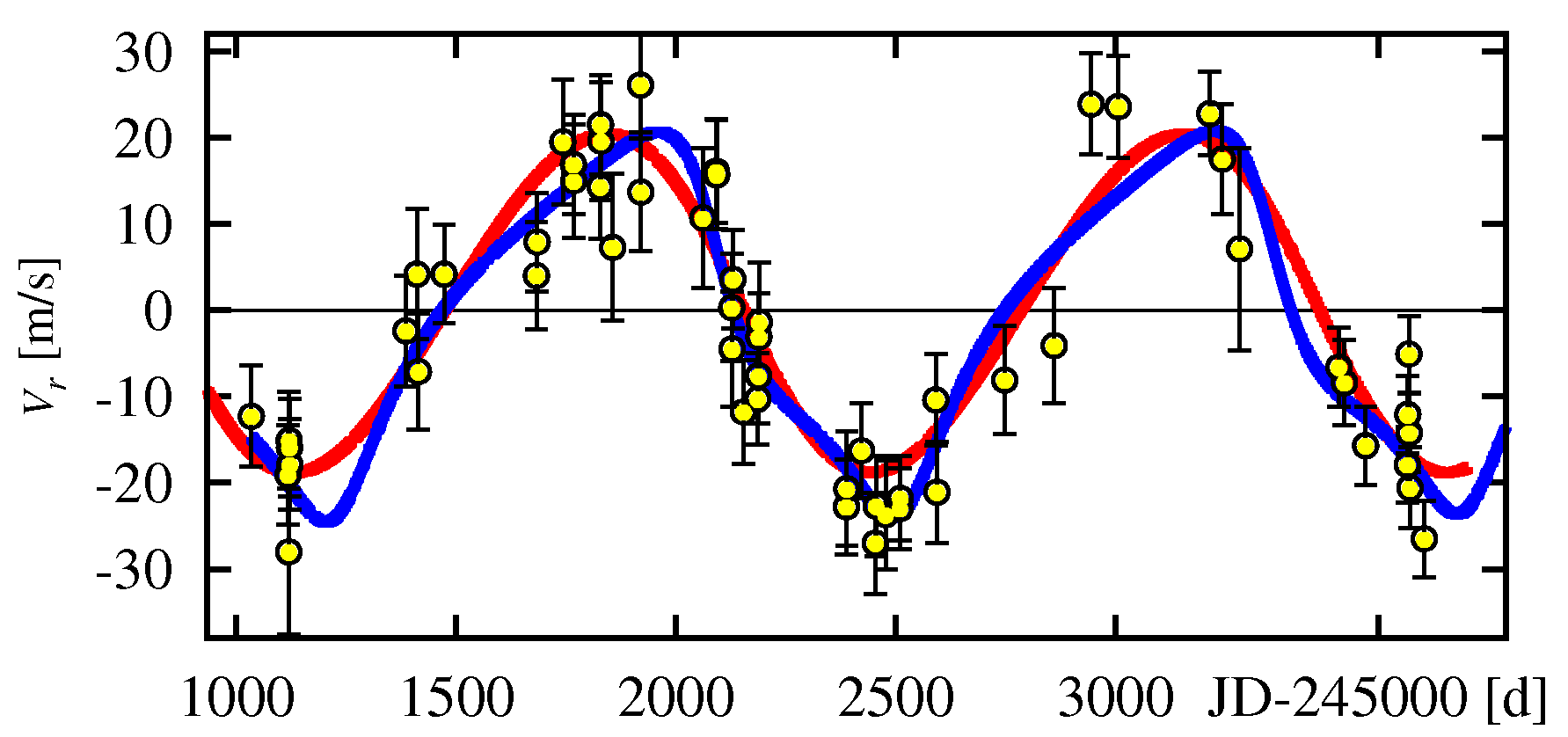}}
\caption{\em 
The RV data of $\tau^1$~Gruis and the synthetic signals. The red curve is for the
the best, single-planet Keplerian fit yielding $\Chi\sim1$ and an rms $\sim6.1$~m/s.
The blue curve (darker one) is for coplanar, edge-on configuration (an rms $\sim5.9$~m/s)
involved in 1b:1c MMR (see the top-left panel in Fig.~\ref{fig:fig5} for the
dynamical map).
}
\label{fig:fig6}
\end{figure}
\else
\begin{figure}
\centerline{\includegraphics[width=80mm]{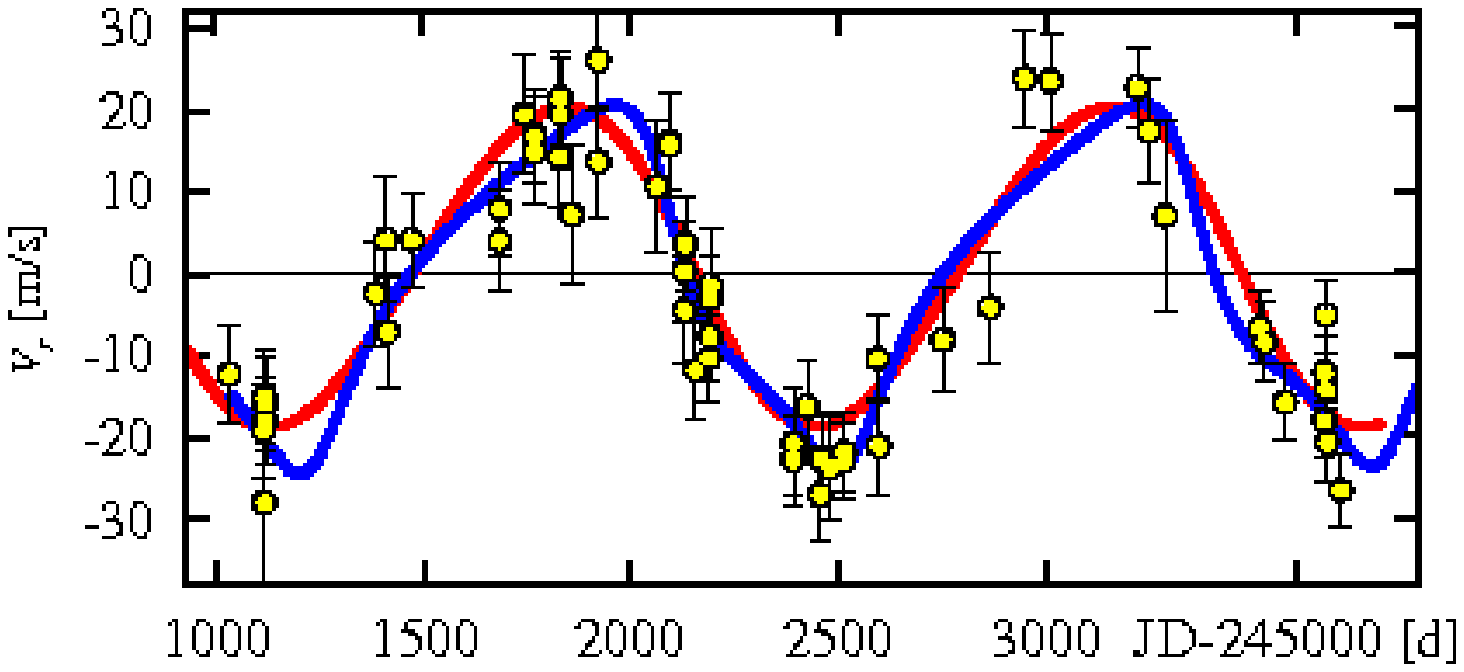}}
\caption{\em 
The RV data of $\tau^1$~Gruis and the synthetic signals. The red curve is for the
the best, single-planet Keplerian fit yielding $\Chi\sim1$ and an rms $\sim6.1$~m/s.
The blue curve (darker one) is for coplanar, edge-on configuration (an rms $\sim5.9$~m/s)
involved in 1b:1c MMR (see the top-left panel in Fig.~\ref{fig:fig5} for the
dynamical map).
}
\label{fig:fig6}
\end{figure}
\fi

The ambiguity of the RV fits implies interesting issues concerning the models of
creation and stability of Earth-like planets interior to the orbits of the
putative Jovian Trojans.  In the $\tau^1$~Gruis, the space interior to the
Jovian planet is ``empty'' as no smaller planets have been yet detected. So we
can try to predict in which regions of the habitable zone ($\sim 1$~AU) smaller
planet could survive. For this purpose we computed dynamical maps for putative
Earth-like masses with initial conditions varied in the $(a_0,e_0)$ plane, and
initial  orbital angles set to $0^{\circ}$. We considered two dynamical
environments: the one with the best-fit Jovian companion in close to circular
orbit and the second one with Trojans in quasi-circular orbits (their elements
are given in Table~1, fit~III). The results are shown in Fig.\ref{fig:fig7}. For
the first configuration, we detect an extended zone of stable motions. Additional
experiments regarding creation of Earth-like planets through coagulation of Mars
and Moon-size protopolanets (see Raymond, 2008 in this volume) performed with the Mercury code  
\citep{Chambers1999}
assures us that such planets emerge easily in that zone. In the case of a
configuration with Trojans, the stable zone shrinks significantly. Moreover, the
creation of Earth-like planets is much more difficult. We found that they could
form only in the zones of relatively stable motions, up to $\sim 0.9$~AU and in
the ``gap'' between the 4:1~MMR and the border of global instability. Yet in
that case, the simulations are very difficult to carry out due to frequent close
encounters between planetesimals and the Jovian planets. 

\ifpdf
\begin{figure}
\centerline{
\hbox{
  \includegraphics[width=52mm]{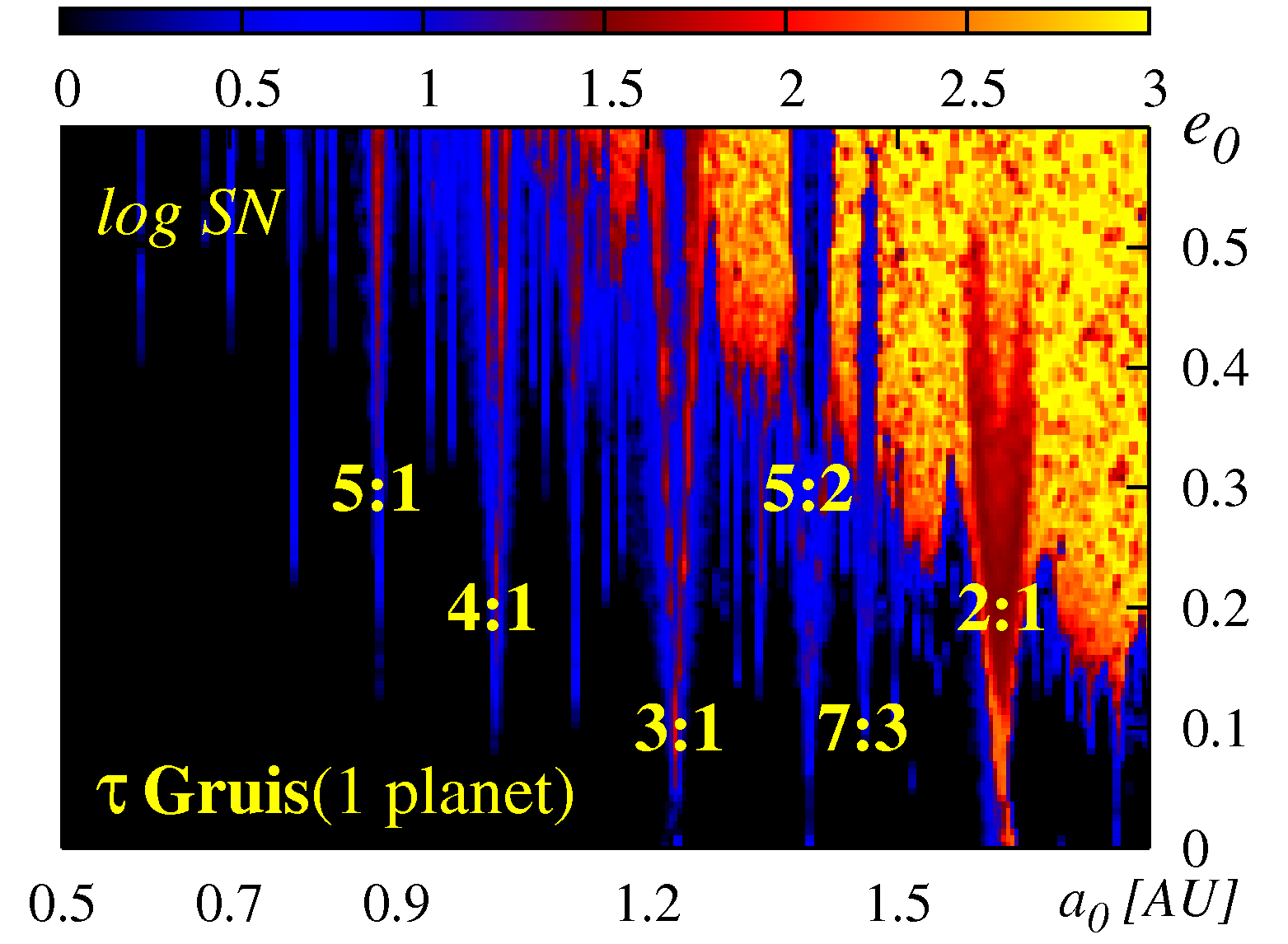}
  \hspace*{5mm}
  \includegraphics[width=52mm]{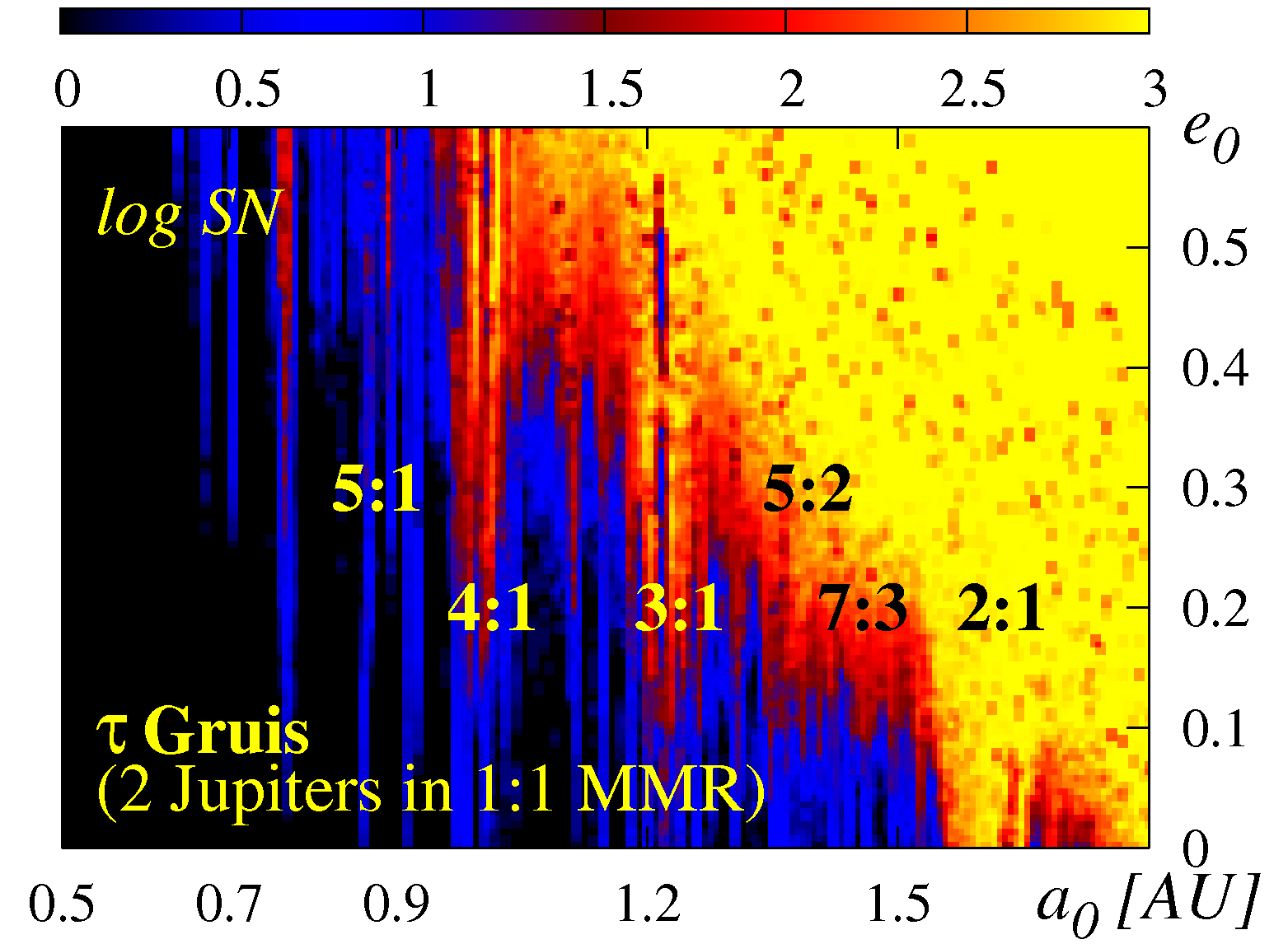}
}
}
\vspace*{0mm}
\centerline{
\hbox{
  \includegraphics[width=52mm]{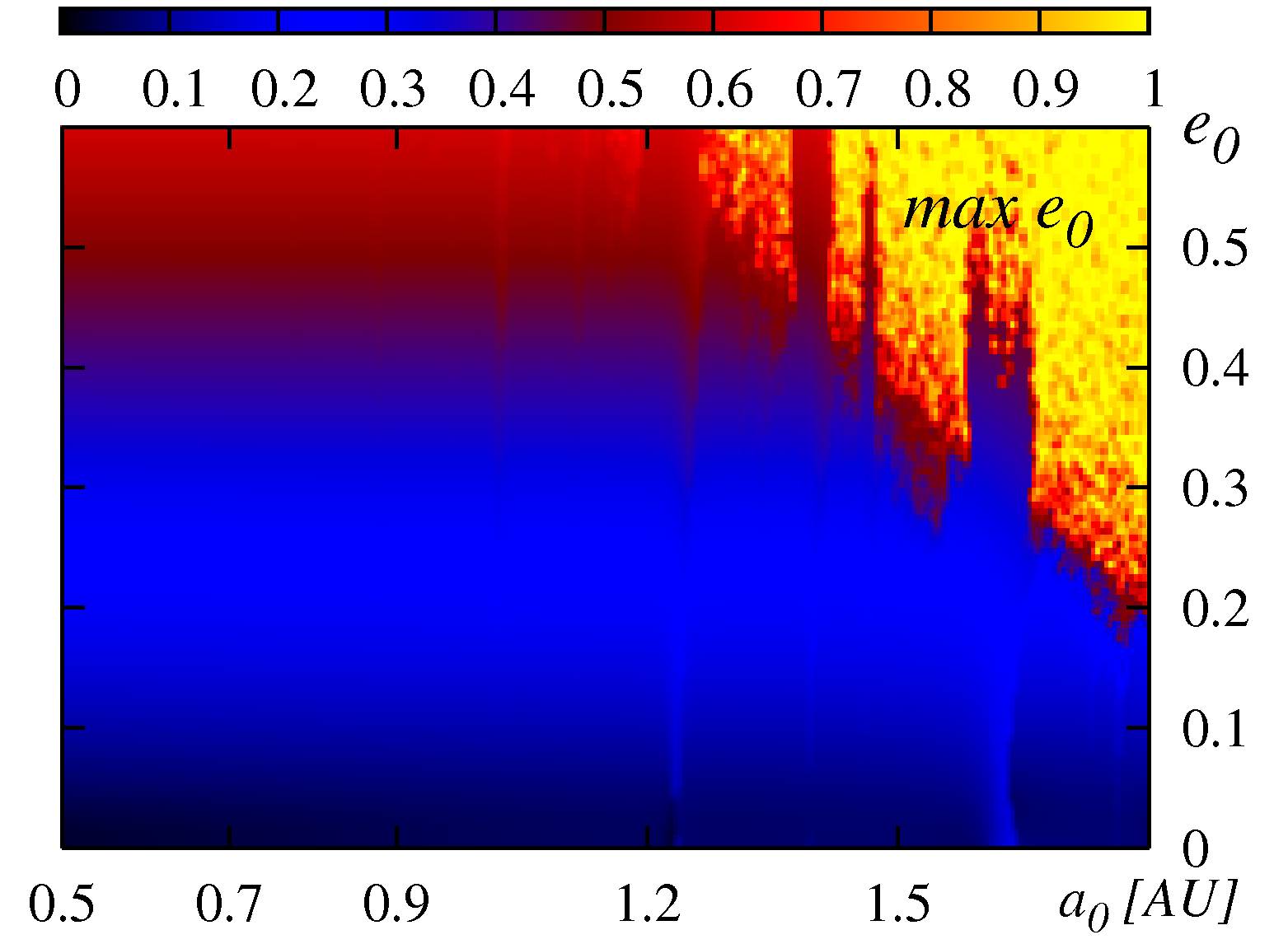}
   \hspace*{5mm}
  \includegraphics[width=52mm]{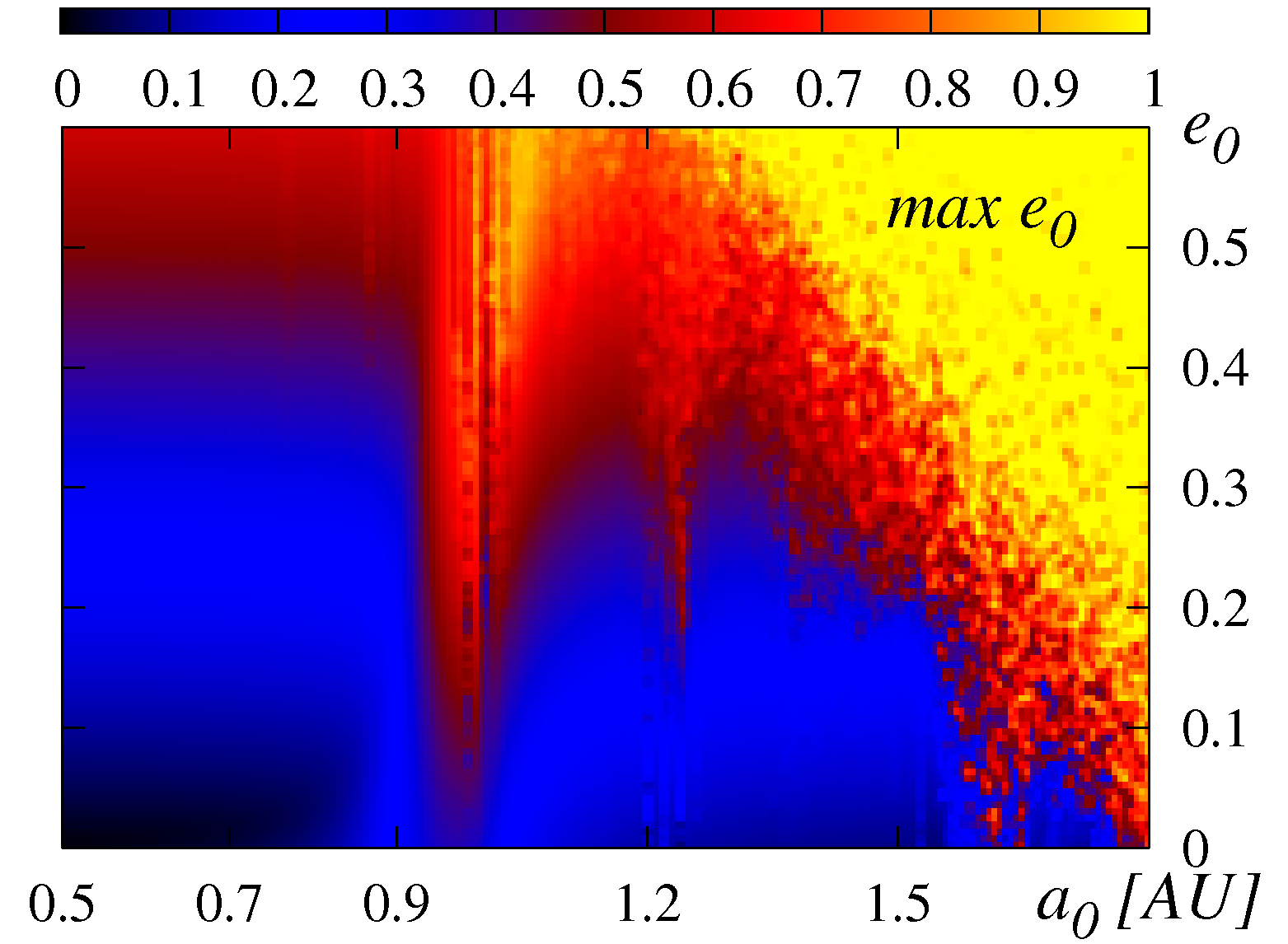}
}
}
\caption{\em 
Dynamical maps for Earth-mass planets in the coplanar $\tau^1$~Gruis system. The
left column is for the stability map for the best fit configuration with one
Jovian planet in quasi-circular orbit ($a \sim2.5$~AU), the right column is for
the systems with Trojans (Table~1, fit~III). Panels in the top row are for the
stability indicator $\log SN$.  Panels in the bottom row are for the $\max e$
indicator (the integration period $\sim 50,000$~yr). Some MMRs between the
Earth-like planet and the Jovian companions are labeled.
}
\label{fig:fig7}
\end{figure}
\else
\begin{figure}
\centerline{
\hbox{
  \includegraphics[width=52mm]{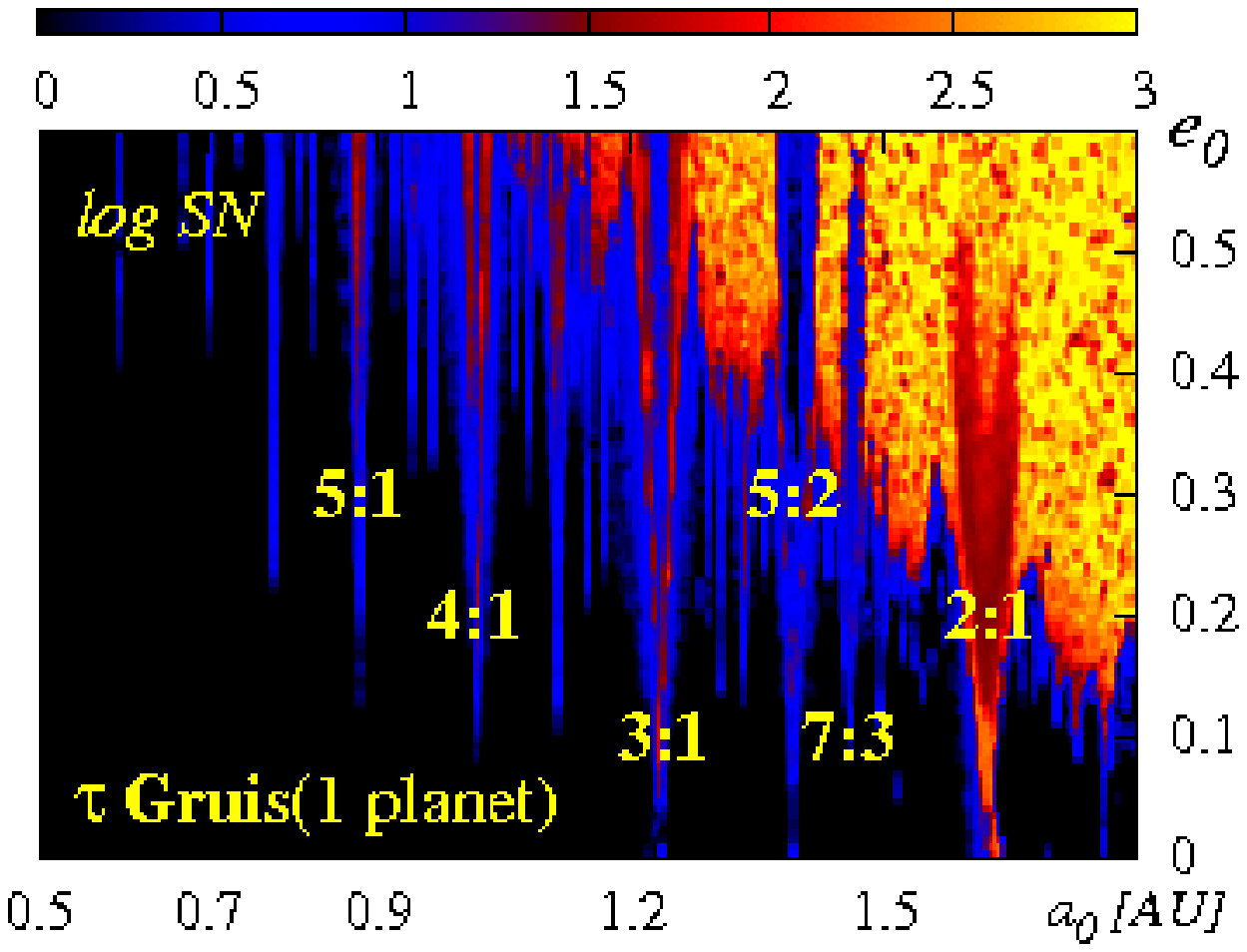}
  \hspace*{5mm}
  \includegraphics[width=52mm]{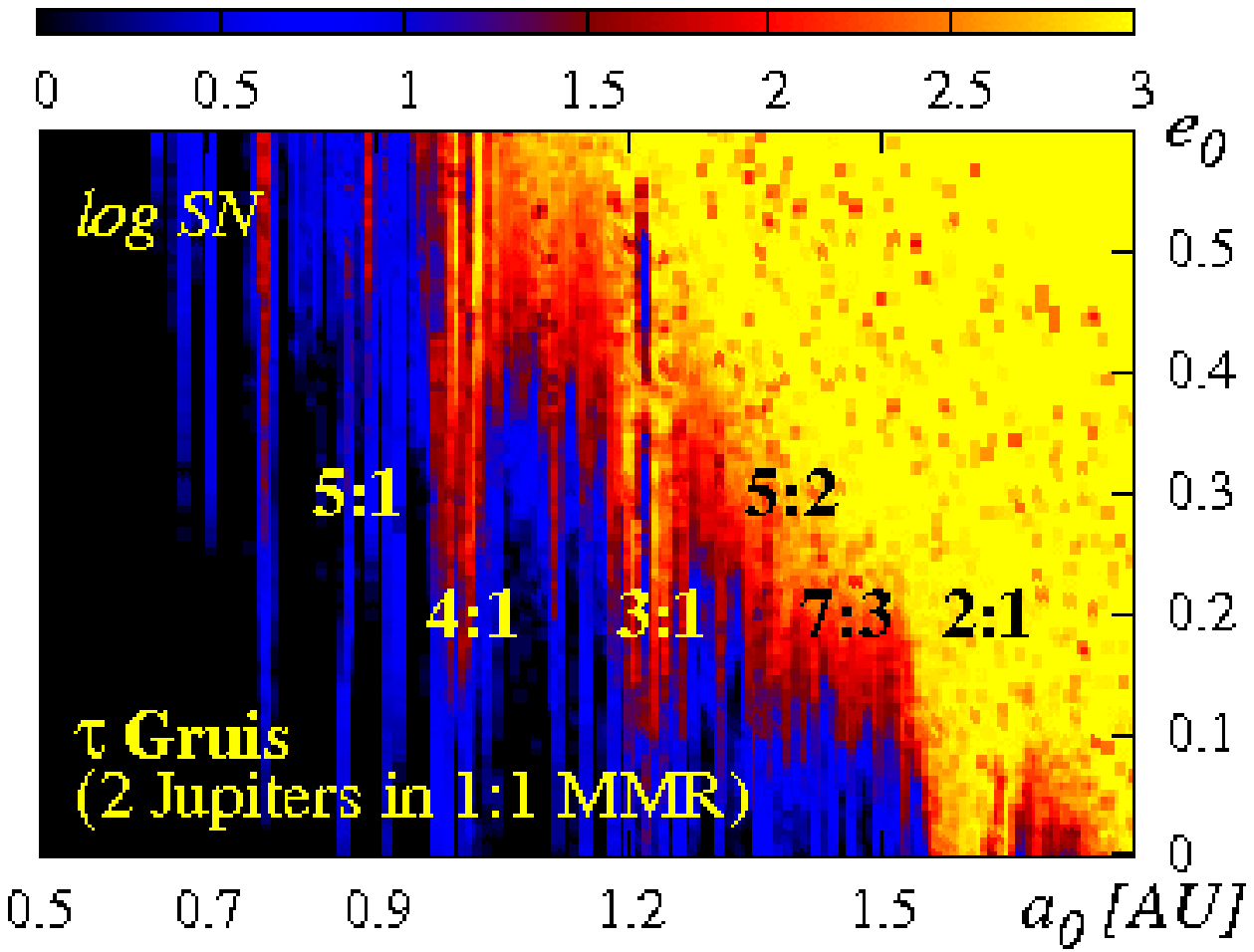}
}
}
\vspace*{0mm}
\centerline{
\hbox{
  \includegraphics[width=52mm]{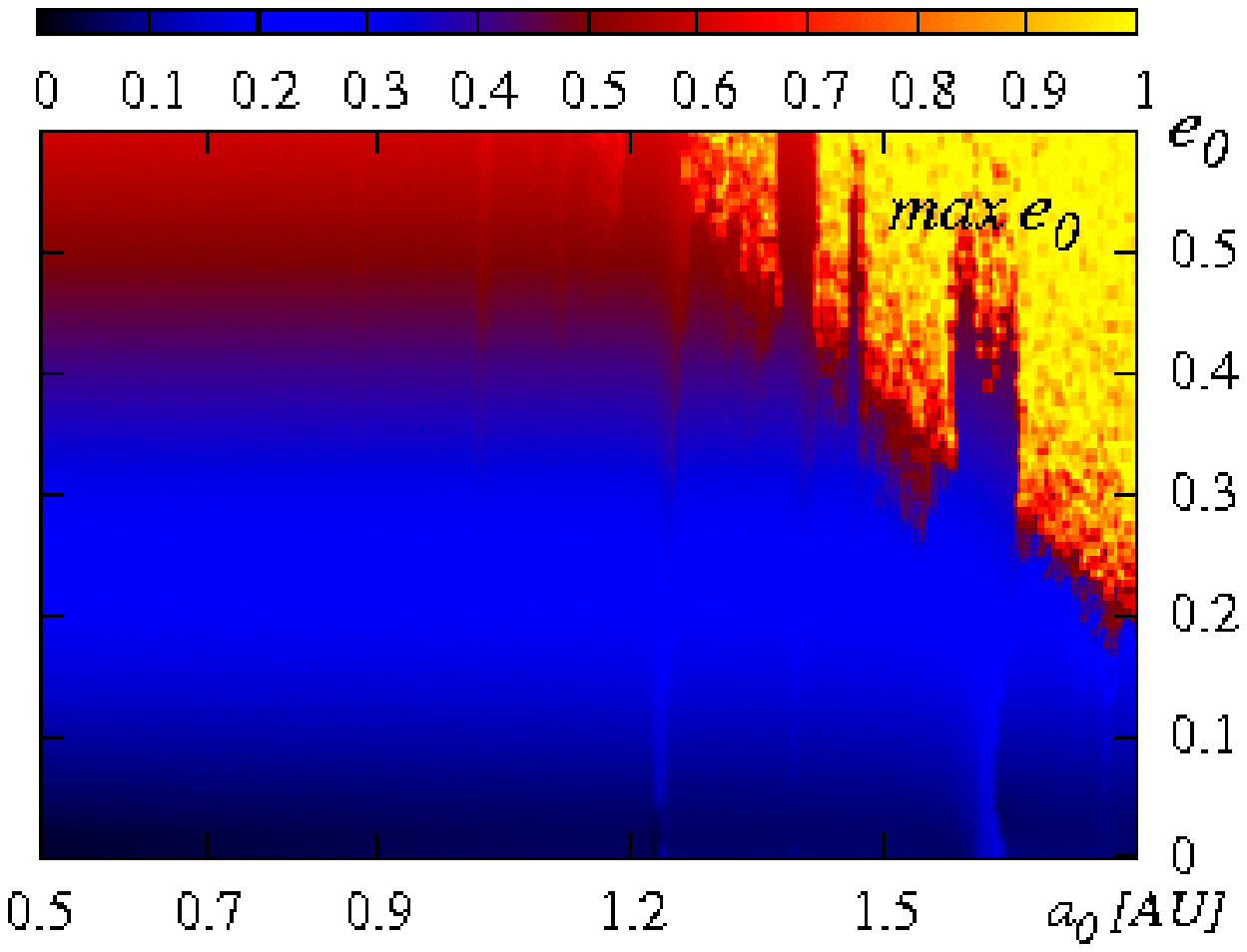}
   \hspace*{5mm}
  \includegraphics[width=52mm]{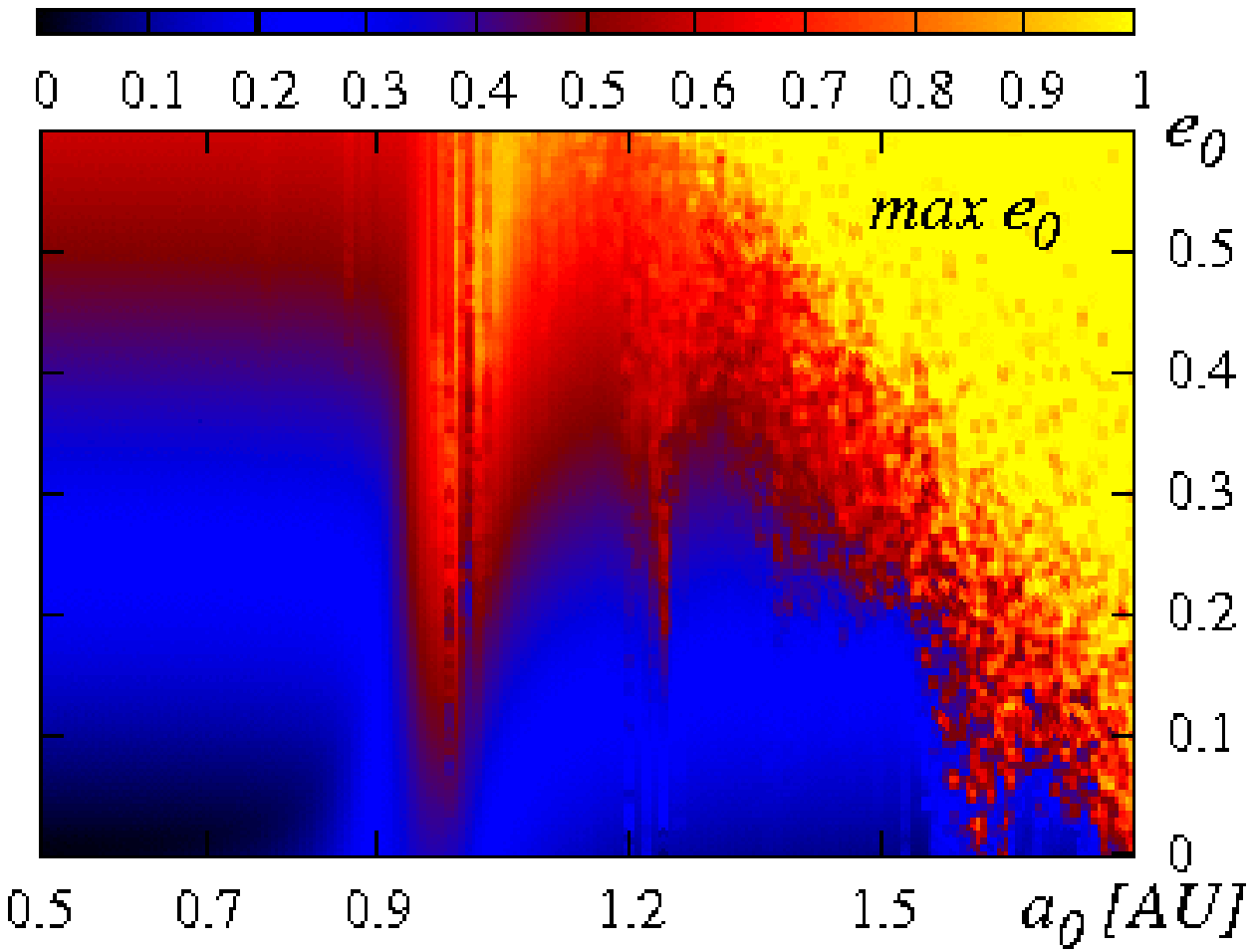}
}
}
\caption{\em 
Dynamical maps for Earth-mass planets in the coplanar $\tau^1$~Gruis system. The
left column is for the stability map for the best fit configuration with one
Jovian planet in quasi-circular orbit ($a \sim2.5$~AU), the right column is for
the systems with Trojans (Table~1, fit~III). Panels in the top row are for the
stability indicator $\log SN$.  Panels in the bottom row are for the $\max e$
indicator (the integration period $\sim 50,000$~yr). Some MMRs between the
Earth-like planet and the Jovian companions are labeled.
}
\label{fig:fig7}
\end{figure}
\fi

\begin{table}
\label{tab:tab1}
\centering
\begin{tabular}{lcccccc}
\hline
& \multicolumn{2}{c}{{\bf HD 155358} (fit {\bf I})} 
& \multicolumn{2}{c}{{\bf HD 155358} (fit {\bf II})} 
& \multicolumn{2}{c}{{\bf $\tau^1$~Gruis} (fit {\bf III})}  \\
Parameter \hspace{1em}
& \ \ planet {\bf b} \ \  & \ \ planet {\bf c} 
& \ \ planet {\bf b} \ \  & \ \ planet {\bf c} 
& \ \ planet {\bf b} \ \  & \ \ planet {\bf c} 
\\
\hline
$m \sin i$ [m$_{\idm{J}}$]
            & 0.827    &    0.490
            & 0.863    &    0.497
            & 0.401     &   0.923     
\\
$a$ [AU]        &  0.623   &  0.875 
                &  0.628     &  1.212  
                &  2.471    &  2.565  
\\
$e$             &   0.121   &     0.743
               &   0.128           &   0.198
               &  0.027     &   0.053   
\\
$\omega$ [deg]      &  131.6    &  86.11
                  &  161.9    &  272.0 
                  &  99.9    & 163.8  
\\
${\cal M}(t_0)$ [deg]
            &   106.32  &  313.7
            &   130.3         & 198.8  
            &   3.5         & 3.6  
\\
$\Chi$         & \multicolumn{2}{c}{1.14} 
               & \multicolumn{2}{c}{1.08}
               & \multicolumn{2}{c}{1.12}
\\
$\sigma_{\idm{j}}$~[m/s] & \multicolumn{2}{c}{5} 
                     & \multicolumn{2}{c}{5} 
                     & \multicolumn{2}{c}{4} 
\\
rms~[m s$^{-1}$]       & \multicolumn{2}{c}{6.31} 
                       & \multicolumn{2}{c}{5.98}
                       & \multicolumn{2}{c}{5.96}
\\
$V_0$ [m s$^{-1}$]         & \multicolumn{2}{c}{12.69}
                            & \multicolumn{2}{c}{10.69}
                            & \multicolumn{2}{c}{ -0.04}
\\
$M_{\star}$ [$M_{\circ}$]   & \multicolumn{2}{c}{0.87}
                            & \multicolumn{2}{c}{0.87}
                            & \multicolumn{2}{c}{1.25}
\\
\hline
\end{tabular}
\caption{\em
The best-fit astro-centric, osculating Keplerian elements of stable, coplanar
and edge-on planetary configurations at the epoch of the respective first
observation.  Original errors of the data are rescaled by
adding the ``jitter'' $\sigma_{\idm{j}}$ in quadrature.
}
\end{table}

\section{Conclusions}
%
In this work we consider some problems related to modeling observations of stars
hosting multi-planet systems. It is well known that the phase space of  such
system has a non-continuous and complex structure with respect to any stability
criterion. Hence, when searching for initial conditions, one has to take into
account the dynamical character of putative planetary configurations. Due to
narrow observational windows, significant measurement errors, stellar jitter and
other uncertainties, the formal best-fits may appear very unstable.
Searching for stable solutions in their neighborhood of the phase space by trial
and error, we should not expect that the results  could be statistically
optimal. Thus, an intuitively natural approach is to eliminate unstable
configurations during the fitting process, through penalizing unstable solutions
with a suitably large value of the $\Chi$ function, or of another measure of the fit
quality. In that way, the stability plays a role of an additional, implicit
observable.  That method is suitable for multi-body systems with Jovian planets
presumably involved in low-order mean motion resonances (MMRs). In particular,
we considered two new examples in which the model of the RV may be non-unique.
The RV of HD~155358 by \cite{Cochran2007} permit a few local minima of $\Chi$
related to different orbital configurations. We also found an example
illustrating the ambiguity of  Keplerian, close to circular single-planet solutions.
The RV data of $\tau^1$~Gruis  could be equally well modeled with coplanar
configurations of Jovian planets involved in 1:1 MMRs. 

Moreover, our fitting method, used mainly for RV data, is quite general and may
be applied to other types of observations as well. As the stability criterion,
one can use the maximal Lyapunov exponent, the most stringent and formal
characteristic of stable/unstable motion. Other suitable indicators like the
maximal eccentricity, the spectral number, or the diffusion rate of fundamental
frequencies may also be applied. These fast indicators help us to search for and
find long-term stable solutions, but also make it possible to efficiently
explore and to visualize the sophisticated and varying structure of the phase
space. We can see the planetary system in its dynamical environment.

Many multi-planet systems are found on the edge of long-term dynamical
stability. It is not clear yet whether this is a general property 
of multi-planet systems, the outcome of poor statistics or 
just the consequence of a bad choice of the RV model. Large eccentricities in
multi-planet systems may ``hide'' other, unknown planets. Yet in that case, the
dynamical modeling of the RV with stability constraints provides valuable
information on the dynamical structure of the putative planetary configurations.
Finding the best fits on the very edge of stable zones may provide good hints 
and motivation to look for alternate models of the RV.

\section*{Acknowledgments}
%
K.~G. thanks the organizers of the IAU~249 symposium for the invitation and 
great hospitality. 
We are grateful to the anonymous referee for
comments that improved the manuscript. Many thanks to Boud Roukema
for corrections of the text.
This work is supported by the Polish
Ministry of Science,  Grant~1P03D~021~29.

\bibliographystyle{plainnat}
\bibliography{biblio}

\end{document}